\documentclass[useAMS,usenatbib]{mnras}

\usepackage{Times}
\usepackage{psfrag}
\usepackage{pspicture}
\usepackage{graphicx}
\usepackage{subfig}
\usepackage{amsmath}
\usepackage{comment}
 \usepackage[flushleft]{threeparttable}

\title[SFGs and their SMBHs over cosmic time]{The growth of typical star-forming galaxies and their super massive black holes across cosmic time since $\bf z\sim2$}
\author[J. Calhau et al.]{Jo\~ao Calhau$^{1,2}$\thanks{E-mail: j.calhau@lancaster.ac.uk}, David Sobral$^{1,2,3}$, Andra Stroe$^4$\thanks{ESO Fellow}, Philip Best$^5$, Ian Smail$^{6}$, Bret Lehmer$^{7}$, \newauthor Chris Harrison$^6$, Alasdair Thomson$^6$\\
$^{1}$ Department of Physics, Lancaster University, Lancaster, LA1 4YB, UK \\
$^2$ Instituto de Astrof\'isica e Ci\^encias do Espa\c{c}o, Universidade de Lisboa, OAL, Tapada da Ajuda, P-1349-018 Lisboa, Portugal\\
$^{3}$ Leiden Observatory, Leiden University, P.O.\ Box 9513, NL-2300 RA Leiden, The Netherlands\\
$^{4}$ European Southern Observatory, Karl-Schwarzschild-Str. 2, 85748, Garching, Germany\\
$^5$ SUPA, Institute for Astronomy, Royal Observatory of Edinburgh, Blackford Hill, Edinburgh EH9 3HJ, UK\\
$^6$ Centre for Extragalactic Astronomy, Department of Physics, Durham University, South Road, Durham DH1 3LE, UK\\
$^7$ Department of Physics, University of Arkansas, 226 Physics Building, 835 West Dickson St., Fayetteville, AR 72701, USA\\
}

\begin{document}

\date{Accepted 2016 September 07 . Received 2016 September 01; in original form 2016 June 24}
\pagerange{\pageref{firstpage}--\pageref{lastpage}} \pubyear{2016}
\maketitle

\label{firstpage}
\begin{abstract}
Understanding galaxy formation and evolution requires studying the interplay between the growth of galaxies and the growth of their black holes across cosmic time. Here we explore a sample of H$\alpha$-selected star-forming galaxies from the HiZELS survey and use the wealth of multi-wavelength data in the COSMOS field (X-rays, far-infrared and radio) to study the relative growth rates between typical galaxies and their central supermassive black holes, from $z=2.23$ to $z=0$. Typical star-forming galaxies at $z\sim1-2$ have black hole accretion rates ($\rm \dot{M}_{\rm BH}$) of 0.001-0.01M$_{\odot}$\,yr$^{-1}$ and star formation rates (SFRs) of $\sim$10-40 M$_{\odot}$\,yr$^{-1}$,  and thus grow their stellar mass much quicker than their black hole mass (3.3$\pm$0.2 orders of magnitude faster). However, $\sim3$\% of the sample (the sources detected directly in the X-rays) show a significantly quicker growth of the black hole mass (up to 1.5 orders of magnitude quicker growth than the typical sources). $\rm \dot{M}_{\rm BH}$ falls from $z=2.23$ to $z=0$, with the decline resembling that of star formation rate density or the typical SFR (SFR$^*$). We find that the average black hole to galaxy growth ($\rm \dot{M}_{\rm BH}$/SFR) is approximately constant for star-forming galaxies in the last 11 Gyrs. The relatively constant $\rm \dot{M}_{\rm BH}$/SFR suggests that these two quantities evolve equivalently through cosmic time and with practically no delay between the two.
\end{abstract}


\section{Introduction}\label{intro}

Understanding how galaxies form and evolve is a very challenging task, as there are a range of complex processes and quantities that need to be taken into account and that usually cannot be studied in isolation, such as gas abundances, dust, supernovae, radiative winds and relativistic jets \citep[e.g.][]{Illustris2015,Eagle2015}. Both the star formation history (SFH; e.g. \citealt{Lilly96,Karim2011,Sobral2013a}) and the black hole accretion history (BHAH; \citealt{Brandt2015})  are strongly influenced by the feedback effects of both star formation (SF) and black hole (BH) accretion, as they affect the ability of the host galaxy to convert molecular gas into stars. For example, an active galactic nucleus (AGN) is the result of the accretion of matter into the central supermassive black hole of a galaxy.  A growing, massive BH releases copious amounts of energy so, provided that there is a strong coupling between radiation and the mechanical output of the BH and surrounding gas, the AGN may be able to disrupt the environment and in principle even quench the SF happening in the host galaxy \citep[e.g.][]{SilkRees1998, Bower2006}. This may happen mainly in two ways: i) radiatively-driven winds and ii) relativistic jets.

Current studies cannot establish whether or not radiatively-driven winds have a significant effect on a galactic scale. Integral field unit (IFU) observations provide evidence for outflowing gas in local Seyferts \citep[e.g.][]{Davies2009, Storchi2010, Muller2011} on scales of $10-100$\,pc. Conversely, spectro-polarimetry of low redshift quasars shows high-velocity outflows close to the accretion disk \citep[e.g.][]{Young2007, GangulyBrotherton}. However, these winds are only observed along the line of sight and there are no direct constraints on the distribution of the outflowing gas, which makes it difficult to get a clear picture of how they affect the galaxy \citep[e.g.][]{Tremonti2007, Dunn2010, Harrison2012}.

Relativistic jets are known to influence gas on a galactic scale, even reaching outside of the dark matter haloes of galaxies and, in addition, interact strongly with virialised hot atmospheres \citep[e.g.][]{Best2005, Nesvadba2006, Nesvadba2007, Nesvadba2008, McNamara2009, McNamara2011}. The accretion of matter into the central black hole leads to the emission of radiation from both the accretion disk and the relativistic jets and thus, in conjunction with star formation processes and gas dynamics, AGN are thought to be responsible for regulating the evolution of galaxies - but it may well be that AGN feedback mostly works as a maintenance mode \citep[e.g.][]{Best2005, Best2006} rather than be responsible for the actual quenching process.

Stellar feedback also plays a major role in regulating star formation. This can happen through extreme events like strong stellar winds or shock waves of supernovae explosions \citep[]{Geach2014}. Typical outflows from star formation involve only small fractions of the molecular gas in Milky Way type galaxies (but are much more important for very low mass galaxies) and thus stellar feedback is generaly considered to be insufficient for the regulation without the contribution of an AGN.

In order to understand how galaxies evolve, it is particularly important to understand how key properties such as the star formation rate (SFR) and the black hole accretion rate ($\rm \dot{M}_{\rm BH}$) in active galactic nuclei (AGN) evolve as a function of cosmic time. This can be done by examining the star formation and black hole accretion histories of galaxies. The latest surveys show that star formation activity peaks at $z\sim2$ \citep[e.g.][]{Sobral2013a,Madau2014} and then declines until today. As for the black hole accretion rates, the peak may happen at slightly lower redshifts than the peak of star formation, but the black hole activity may also decline more rapidly from $z\sim1$ to 0 \citep[e.g.][]{Aird2010}. However, studies taking into account the bolometric luminosity functions of AGN \citep[e.g.][]{Delvecchio2014} show that black hole accretion tracks the evolution of SF more closely, peaking at $z\sim2$.

Most studies on the evolution of SF and BH accretion tend to focus on AGN selected samples. \cite{Stanley2015}, for example, found that while there is a strong evolution of the average SFR with redshift, the relation between SFR and AGN luminosity seems relatively flat for all redshifts. The authors interpreted this as being due to the effect of short time-scale variations in the mass accretion rates, which might erase any relation that might exist between the SFR and AGN luminosity. Nevertheless, there are also studies with star-forming selected samples: \cite{Delvecchio2015} analysed the relation of AGN accretion and SFR for star-forming galaxies up to $z\sim2.5$ and found that the ratio between the $\rm \dot{M}_{\rm BH}$ and the SFR evolves slightly with redshift, and has a lower value compared to what one would need to obtain the local M$\rm _{BH}$-M$\rm _{Bulge}$ relation. \cite{Lehmer2013} also investigated the $\rm \dot{M}_{\rm BH}$/SFR ratio using galaxy samples from both the field and a high-density structure (super-cluster of QSO from the 2QZ survey) at $z\sim2.23$. \cite{Lehmer2013} found that H$\alpha$ emitting galaxies in this structure have a relatively high fraction of AGN activity, leading to average $\rm \dot{M}_{\rm BH}$/SFR which are closer to what is typically measured for AGN. For more typical ``field'' H$\alpha$ emitters, the $\rm \dot{M}_{\rm BH}$/SFR was found to be typically an order of magnitude lower than for AGN and for H$\alpha$ emitters in the higher density region at $z\sim2$. These results suggest that SF galaxies are generally situated below the local relation (at least at redshifts of $z\sim2$) and that the activity of the AGN causes the ratio to rise high enough so that the galaxies approach a growth mode that could easily result in the observed local relation. However, much is still unknown, for typical, star-formation selected samples, regarding the relative growth of the black hole and the host galaxies, and particularly how such relative growth may vary with time, from the peak of the star formation history, at $z\sim2.5$ to $z\sim0$.

In this paper we explore a sample of ``typical'' star-forming galaxies from HiZELS in the COSMOS field, selected in four different redshift slices in a self-consistent, homogenous way. We explore the wealth and variety of exquisite data in the COSMOS field to study the relative growth between the central black holes and their host galaxies, and how that varies across cosmic time. This paper is organised as follows: Section 2 presents the data and sample. Section 3 provides an overview of our selection of potential AGNs. Section 4 presents our stacking analysis in different bands. Section 5 presents the results: the relative supermassive black hole/galaxy growth and in section 6 we present the conclusions. In this paper, we use a Chabrier IMF \citep[]{Chabrier2003} and the following cosmology: H$_0$=70 km\,s\,$^{-1}$\,Mpc$^{-1}$, $\Omega_M$=0.3 and $\Omega_{\Lambda}$=0.7.

\section{Data and sample}\label{sample_data}

\subsection{Data: X-rays, radio \& FIR} \label{COSMOS_data}

\subsubsection{X-rays: C-COSMOS}\label{x_ray_DATA}

The \textit{Chandra} Cosmos Survey \citep[C-COSMOS;][]{Elvis2009, Puccetti2009} imaged the COSMOS field \citep[]{Scoville2007} with an effective exposure time of $\sim$180\,ks and a resolution of $0.5''$. The limiting source detection depths are $1.9\times10^{-16}$\,erg\,s$^{-1}$\,cm$^{-2}$ in the soft band (0.5-2\,keV), $7.3\times$10$^{-16}$\,erg\,s$^{-1}$\,cm$^{-2}$ in the hard band (2-10 keV), and $5.7\times10^{-16}$\,erg\,s$^{-1}$\,cm$^{-2}$ in the full band (0.5-10\,keV). The data allows us to track X-ray emission from processes like Bremsstrahlung and inverse Compton scattering, and thus to identify which sources are AGN based on their X-ray emission. C-COSMOS only covers the relatively central area of COSMOS (0.9\,deg$^2$), and thus we restrict our analysis to that region.

\subsubsection{Radio: VLA-COSMOS}

The VLA-COSMOS Survey \citep[]{Schinnerer2004, Schinnerer2007, Bondi2008} used the National Radio Astronomy Observatory's Very Large Array (VLA) to conduct deep ($\sigma_{1.4}\sim10$\,$\mu$Jy/beam), wide-field imaging with $\approx1.5''$ resolution at 1.4\,GHz continuum of the 2 square-degree COSMOS field. With this band, we track the radio emission of AGN via synchrotron radiation from SMBH relativistic jets and estimate SFRs from the synchrotron radiation due to supernovae explosions \citep[]{Schmitt2006}.

\subsubsection{Far-infrared: Herschel}

COSMOS was imaged with the \textit{Herschel} telescope as part of the \textit{Herschel} Multi-tiered Extragalactic Survey, HerMES \citep{Oliver2012}. HerMES is a legacy program that mapped 380\,deg$^{2}$ of the sky - \textit{Herschel}-SPIRE \citep[250\,$\mu$m, 350\,$\mu$m and 500\,$\mu$m, with a PSF FWHM of 18.1$''$, 24.9$''$ and 36.6$''$, respectively;][]{Griffin2010}. We aditionally make use of the \textit{Herschel} PACS Evolutionary Probe program \citep[PEP: 100\,$\mu$m and 160\,$\mu$m, with PSFs of 7.2$''$ and 12$''$; ][]{Lulz2011} and the observations of the Submillimiter Common-User Bolometer Array 2 (SCUBA2) on the James Clerk Maxwell Telescope, at 850$\mu$m, for the COSMOS Legacy Survey \citep{Geach2013, Geach2016}. These bands cover the peak of the redshifted thermal spectral energy distribution from interstellar dust for galaxies in the redshift range ($z \sim0.4-2.2$) for the entire COSMOS field. The bands therefore capture optical and UV radiation that has been absorbed and re-emitted by dust.

\subsection{The sample of H$\alpha$ emitters at $\bf z=0.4-2.23$}\label{Hizels_survey}

\begin{figure}
\centering
\includegraphics[width=8.3cm]{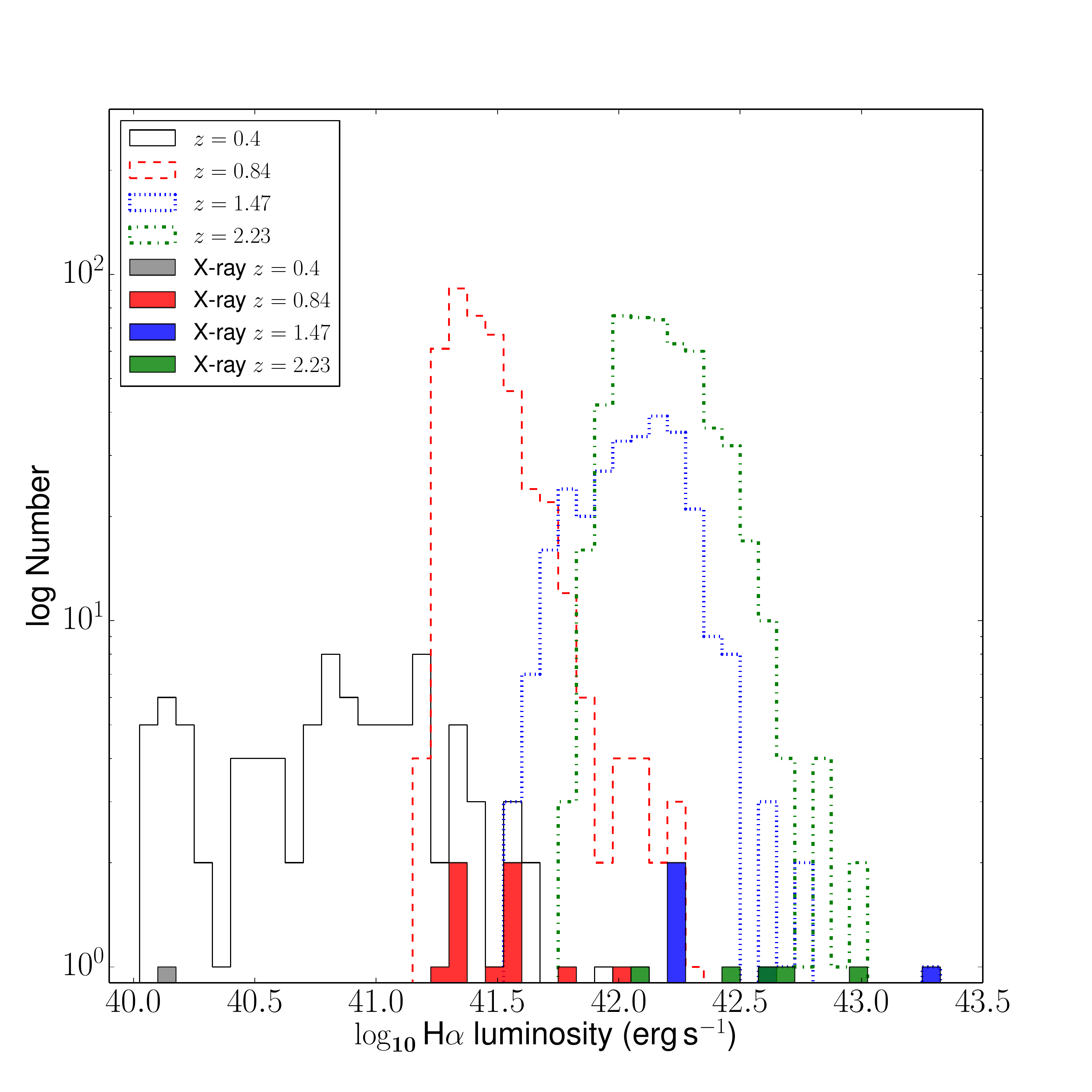}
\caption{H$\alpha$ luminosity distribution of the sample of H$\alpha$ emitters that are used in this paper (after the aplication of a stellar mass cut, see \S \ref{Hizels_survey}) and those with individually detected X-ray emission (filled histograms). X-ray detected H$\alpha$ emitters have `typical' to high H$\alpha$ luminosities. Note that the $z=0.40$ sample covers a much smaller volume than those at higher redshift, thus missing luminous and rarer sources.}
\label{fig:subfigures}
\end{figure}

The High Redshift Emission Line Survey \citep[HiZELS;][]{Geach2008, Best2010, Sobral2009a, Sobral2009b, Sobral2012, Sobral2013a} has surveyed some of the best-studied extragalactic fields for H$\alpha$ emitters at various narrow redshift ranges, from $z=0.4$ to $z=2.23$ \citep[see][]{Sobral2013a}. HiZELS used a set of narrow-band filters in the near-infrared $J$, $H$ and $K$ bands and the Wide Field CAMera \citep[WFCAM, ][]{Casali2007} on the United Kingdom Infrared Telescope (UKIRT), coupled with a filter in the $z'$ band \citep[NB921;][]{Sobral2012, Sobral2013a} mounted on Suprime-cam on the Subaru telescope, to cover roughly 5\,deg$^2$ of extragalactic sky. While it is true that using only H$\alpha$ as a tracer for star formation may cause us to miss obscured star formation, the use of bluer bands for the detection of star-forming galaxies (UV or bluer emission lines) would result in missing a much more significant part of the population. In addition, \cite{Oteo2015} showed that an H$\alpha$ selection is able to recover $\sim$100\% of star-forming galaxies (including the most dusty ones), and \textit{Herschel} is then ideal to recover the full SFRs of such highly obscured galaxies \citep[e.g.][]{Ibar2013}.
Although HiZELS covers various fields, in this work we focus only on the COSMOS field due to the availability of deep data from the \textit{Chandra} Observatory, on which we rely in order to measure the X-ray luminosities in our samples. HiZELS obtained large samples of H$\alpha$ selected galaxies at redshifts $z=0.4$, $z=0.84$, $z=1.47$ and $z=2.23$ in the COSMOS and UDS fields \citep[][]{Sobral2013a}. The H$\alpha$ emitters were selected using a combination of broad-band colours (colour-colour selections) and photometric redshifts. Spectroscopically confirmed sources are included in the sample and the sources confirmed to be other emission line emitters are removed. We refer the interested reader to \cite{Sobral2013a} for the detailed explanation of the process of selection for the H$\alpha$ emitters. Furthermore, we note that while the HiZELS sample at $z=0.4$ (obtained with the Subaru telescope) probes down to significantly lower H$\alpha$ luminosities and stellar masses \citep[see][]{Sobral2014} than those at higher redshift, it also covers a significantly smaller volume, and thus misses massive, bright sources (see Figure \ref{fig:subfigures}). In an attempt to make the $z=0.4$ sample more comparable to those at higher redshift, we apply a mass cut of M\,$>10^9$\,M$_{\odot}$. As we will rely on \textit{Chandra} data for deep X-ray data (\S\ref{x_ray_DATA}), we also need to restrict our analysis to the area in COSMOS with deep \textit{Chandra} coverage. Thus, our final sample is composed of 35, 224, 137 and 276 H$\alpha$ emitters at $z=0.40$, $z=0.84$, $z=1.47$ and $z=2.23$. These are the sources restricted by \textit{Chandra} coverage but include both the ones detected in the C-COSMOS survey and the ones without detectable X-ray emission. We present the distribution of H$\alpha$ (observed luminosities) in Figure \ref{fig:subfigures}.

%
%
%
%

\section{AGN selection}\label{AGN_selection}

\subsection{X-ray detections}

X-rays are one of the best ways to search for AGN. As matter falls into the black hole, it heats up, leading to the emission of radiation in the X-rays through inverse-Compton scattering of UV emission comming from the accretion disk. As the X-ray luminosity is expected to scale with the accretion rate, we can use X-ray luminosities to not only identify AGN, but also to obtain an estimate of the SMBH growth rates.

%
%

We cross-correlate our sample of H$\alpha$ emitters with the \textit{Chandra} X-ray catalogue with a  1$''$ matching radius, in order to find which of our sources are directly detected in the X-rays and thus likely AGN. We find one direct detection at $z=0.4$ ($2.9\pm1.7$\% of the total sample at this redshift), seven at $z=0.84$ ($3.1\pm1.8$\%), four at $z=1.47$ ($2.9\pm1.7$\%) and five at $z=2.23$ ($1.8\pm1.3$\%) in the C-COSMOS catalogue. The results are presented in Table \ref{table:detsourc}. The directly detected sources possess X-ray luminosities of the order of $\geq$10$^{42}$\,erg\,s$^{-1}$, which are typical of the luminosities expected from AGN in this band. Our results are consistent with a non-evolving fraction of X-ray AGN within H$\alpha$ selected samples over the last 11\,Gyrs of cosmic time (since $z\sim2.2$), although we have low number statistics. In Figure \ref{fig:subfigures} we present the H$\alpha$ luminosity distribution of the directly detected AGN, finding that they have preferentially higher than average H$\alpha$ luminosities, raising the possibility that our sources might be contaminated in the H$\alpha$ by AGN. 

%
%
%
\begin{table}
\centering
\caption[]{The luminosity in the X-rays and central $\rm \dot{M}_{\rm BH}$ for the sources directly detected by the C-COSMOS survey (all sources directly detected have luminosities higher than $10^{41}$\,erg\,s$^{-1}$).\\
  * - These sources were taken directly from the tables of the HiZELS survey. In order to get the HiZELS designation for each galaxy, one should add "HiZELS-COSMOS-NB\# DTC" to the beginning of the source's name, where \# stands for the number or letter identifying the filter.}
\begin{tabular}{@{}cccc@{}}
\hline
  Source ID* & Redshift & $\log_{10}$ L$_X$& $\dot{M}_{\rm BH}$\\
   (S13)   &   & [erg\,s$^{-1}$] & [M$_{\odot}$\,yr$^{-1}$]\\
\hline
S12-93079 & 0.40 & 41.97$\pm$0.09  & 0.003$\pm$0.0008\\
S12-22675 & 0.84 & 43.32$\pm$0.04  & 0.074$\pm$0.008\\
S12-33061 & 0.84 & 43.77$\pm$0.03 & 0.207$\pm$0.016\\
S12-26956 & 0.84 & 43.89$\pm$0.03& 0.273$\pm$0.02\\
S12-11275 & 0.84 & 42.76$\pm$0.09& 0.02$\pm$0.004\\
S12-6454 & 0.84 & 42.85$\pm$0.07 & 0.024$\pm$0.005\\ 
S12-4541 & 0.84 & 42.96$\pm$0.08& 0.032$\pm$0.007\\
S12-2436 & 0.84 & 42.69$\pm$0.14& 0.017$\pm$0.006\\
S12-23041 & 1.47 & 43.93$\pm$0.04& 0.3$\pm$0.032\\
S12-19279 & 1.47 & 44.88$\pm$0.01 & 2.69$\pm$0.074\\
S12-20593 & 1.47 & 43.40$\pm$0.07 & 0.087$\pm$0.016\\
S12-44372 & 1.47 & 42.96$\pm$0.14 & 0.032$\pm$0.013\\
S12B-1528 & 2.23 & 43.67$\pm$0.08 & 0.16$\pm$0.033\\
S12B-1073 & 2.23 & 43.48$\pm$0.11 & 0.106$\pm$0.032\\
S12B-9274 & 2.23 & 43.66$\pm$0.10 & 0.098$\pm$0.042\\
S12B-1139 & 2.23 & 43.38$\pm$0.14 & 0.085$\pm$0.032\\
S12B-2306 & 2.23 & 43.45$\pm$0.13& 0.1$\pm$0.035\\
\hline
\label{table:detsourc}
\end{tabular}
\end{table}

\subsection{Radio detections}\label{section:Radio}


We cross correlated the VLA-COSMOS deep catalogue with our H$\alpha$ emitters. Our match between the VLA-COSMOS and our sources resulted in: i) one source is detected at $z=0.4$ ($2.9\pm1.7$\%), 11 radio sources for $z=0.84$ ($4.9\pm2.2$\%), 7 sources for $z=1.47$ ($5.1\pm2.3$\%) and 9 for $z=2.23$ ($3.3\pm1.8$\%). We estimated the radio luminosities by using:
\begin{equation}
\rm L_{1.4GHz}=4 \pi {d_L}^2 S_{1.4GHz} 10^{-33} (1+z)^{\alpha -1} \, (W Hz^{-1}),
\end{equation}
where $\rm d_L$ is the luminosity distance (in cm), $\rm S_{1.4GHz}$ is the flux density in mJy and $\alpha$ is the radio spectral index - assumed to be 0.8, the characteristic spectral index of synchrotron radiation. 0.8 is a good average value for SF-dominated galaxies \citep[e.g. ][]{Thomson2014}, although it is not clear if this value is the best choice if the sample contains a large quantity of AGN. Our SF-selected sample should not have too many AGN (see Table \ref{table:AGN}) so $\rm \alpha = 0.8$ should be appropriate.  Within our H$\alpha$ emitters, the radio sources have radio luminosities of the order of $\sim10^{21}$\,W\,Hz$^{-1}$ at $z=0.40$, $\sim10^{23}$\,W\,Hz$^{-1}$ for $z=0.84-2.23$. It is possible for radio detect up emission from a population of supernova remnants as well. However the emission from these sources have lower luminosity than the AGN we are tracing and should not contaminate the measurements.

\section{Stacking analysis: \textbf{$\rm \dot{M}_{\rm BH}$} and SFR}

\subsection{Radio stacking: SFR}

\begin{figure}
	\centering
	\includegraphics[width=7.4cm]{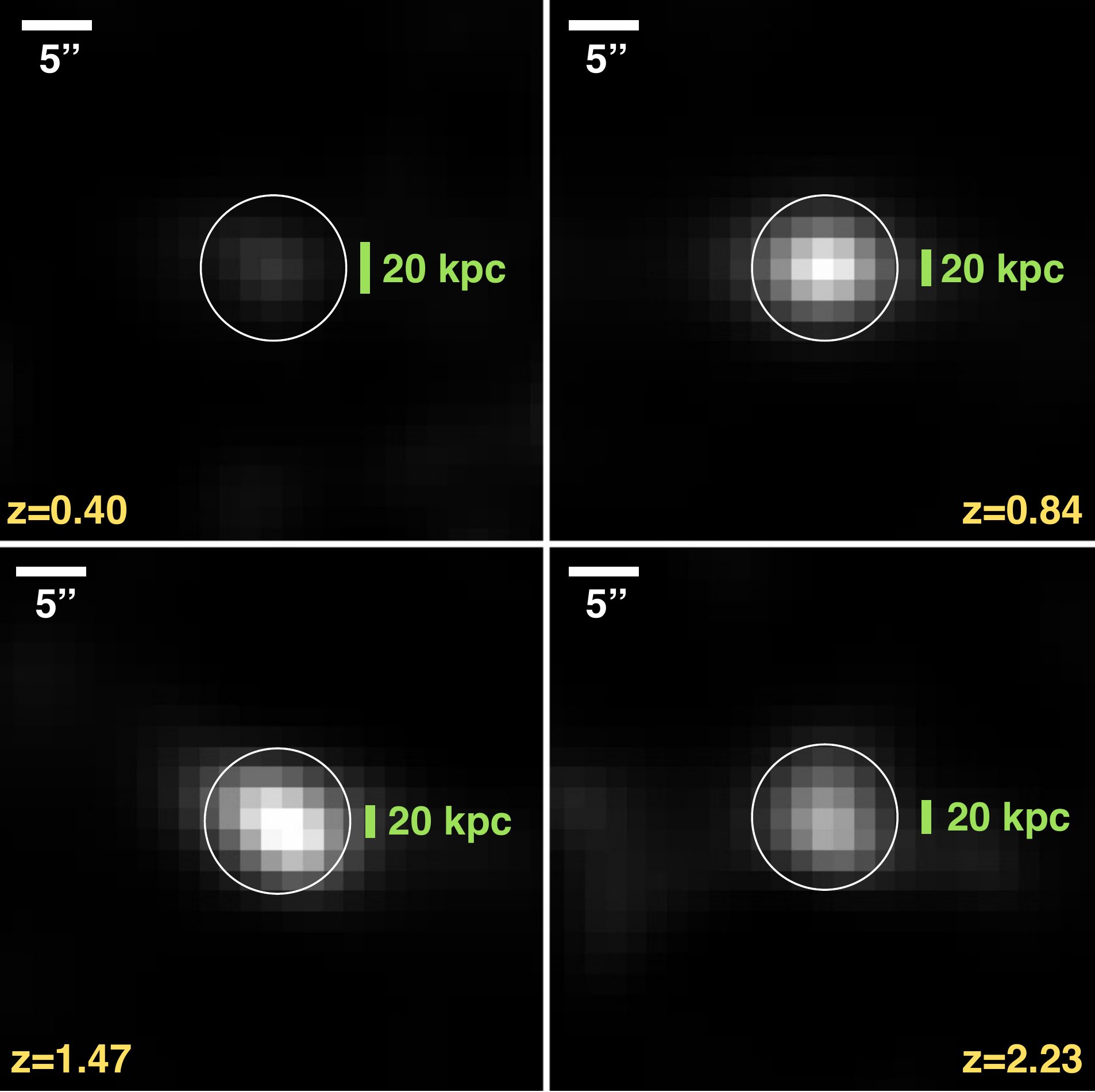}
	\caption{Stacking in the radio (1.4\,Ghz) for our non-radio AGN sources, at each redshift. We find strong detections at every redshift with luminosities of $\approx10^{21-23}$\,W\,Hz$^{-1}$, corresponding to SFRs of $\sim1.5-63$\,M$_{\odot}$. The images were smoothed for easier inspection.}
	\label{fig:RadioStack}
\end{figure}

After rejecting all strong radio sources within our H$\alpha$ selected samples, we can stack the remaining sources, and use radio luminosities as a dust-free star-formation indicator (although some contribution of lower luminosity AGN will still be present, thus likely biasing results towards high star formation rates). We follow the same stacking procedure as for our X-ray stacking (see section \ref{X-ray stacking}) and find high S/N detections of our mean radio stacks in every redshift (see Figure \ref{fig:RadioStack}). We find radio luminosities of $4.6\times10^{21}$\,W\,Hz$^{-1}$, $3.3\times10^{22}$\,W\,Hz$^{-1}$, $2.0\times10^{23}$\,W\,Hz$^{-1}$ and $1.0\times10^{23}$\,W\,Hz$^{-1}$ for $z=0.4$, $0.84$, $1.47$ and $2.23$ respectively.

To convert the luminosities to SFR, we adopted the conversion determined by \cite{Yun2001} converted to a Chabrier IMF \citep[e.g.][]{Karim2011}:
\begin{equation}
\rm SFR_{1.4GHz}=3.18 \times 10^{-22}L_{1.4GHz}\,\,(M_{\odot}\,yr^{-1}).
\end{equation}

The conversion is suitable for radio luminosities up to, and including, 10$^{24}$\,W\,Hz$^{-1}$ and thus expected to yield reasonable results. We find SFRs of $\approx1.5$\,M$_\odot$\,yr$^{-1}$, $10.5$\,M$_{\odot}$\,yr$^{-1}$, $62$\,M$_{\odot}$\,yr$^{-1}$ and $21$\,M$_{\odot}$\,yr$^{-1}$ at $z=0.4$, $0.84$, $1.47$ and $2.23$ respectively.

%
%
%
%

\subsection{FIR stacking: SFRs}\label{section:FIR stacking}

When estimating the SFR, it is important to make sure that there is no contamination to the luminosities by the activity of the AGN. FIR emission from the cold dust \citep[rest frame 40-500$\mu$m][]{Rowan-Robinson1995, Schweitzer2006, Netzer2007} should have little to no such contamination.

Cross-correlating our sample with the HerMES catalogue with a 1'' matching radius resulted in 2 sources being directly detected for $z=0.4$, 10 for $z=0.84$, 5 for $z=1.47$ and 7 galaxies directly detected for $z=2.23$  \citep[See also][]{Ibar2013, Oteo2015}. As expected, most of the sample, made of much more ``typical" star-forming galaxies, is below the depth of \textit{Herschel}, or SCUBA-2, in COSMOS. However, by the means of stacking, one can reach much lower flux limits, and thus detect the mean star-forming galaxy at each redshift.
In order to obtain the necessary SFRs we make use of the results achieved by \cite{Alasdair}. The stacks were obtained through mean statistics accounting for background emission and confusion noise. Aperture corrections were applied for the PACS 100$\mu$m and 160$\mu$m bands, as specified in the PACS PEP release notes. In the SPIRE 250, 350 and 500$\mu$m, the fluxes were taken from the peak value in each stack.
The IR luminosities were then estimated by fitting modified black-body (grey-body) templates to the data points and integrating the best fit between 100 and 850$\mu$m (see \ref{fig:SED}). We refer the interested reader to \cite{Alasdair} for the description of the complete procedure.

We use the total FIR luminosity to compute SFRs (Chabrier IMF) by using:

\begin{equation}
\rm SFR=L_{IR} \times 2.5 \times 10^{-44} \, \,  (M_{\odot} yr^{-1}).
\end{equation}

This translates to a SFR ranging from $2-38$ M$_{\odot}$ yr$^{-1}$ at $z=0.4-2.23$ (see Table \ref{table:stacsourc}).

\subsection{X-ray stacking}\label{X-ray stacking}

%
%
%
\begin{figure}
	\centering
	\includegraphics[width=8.4cm]{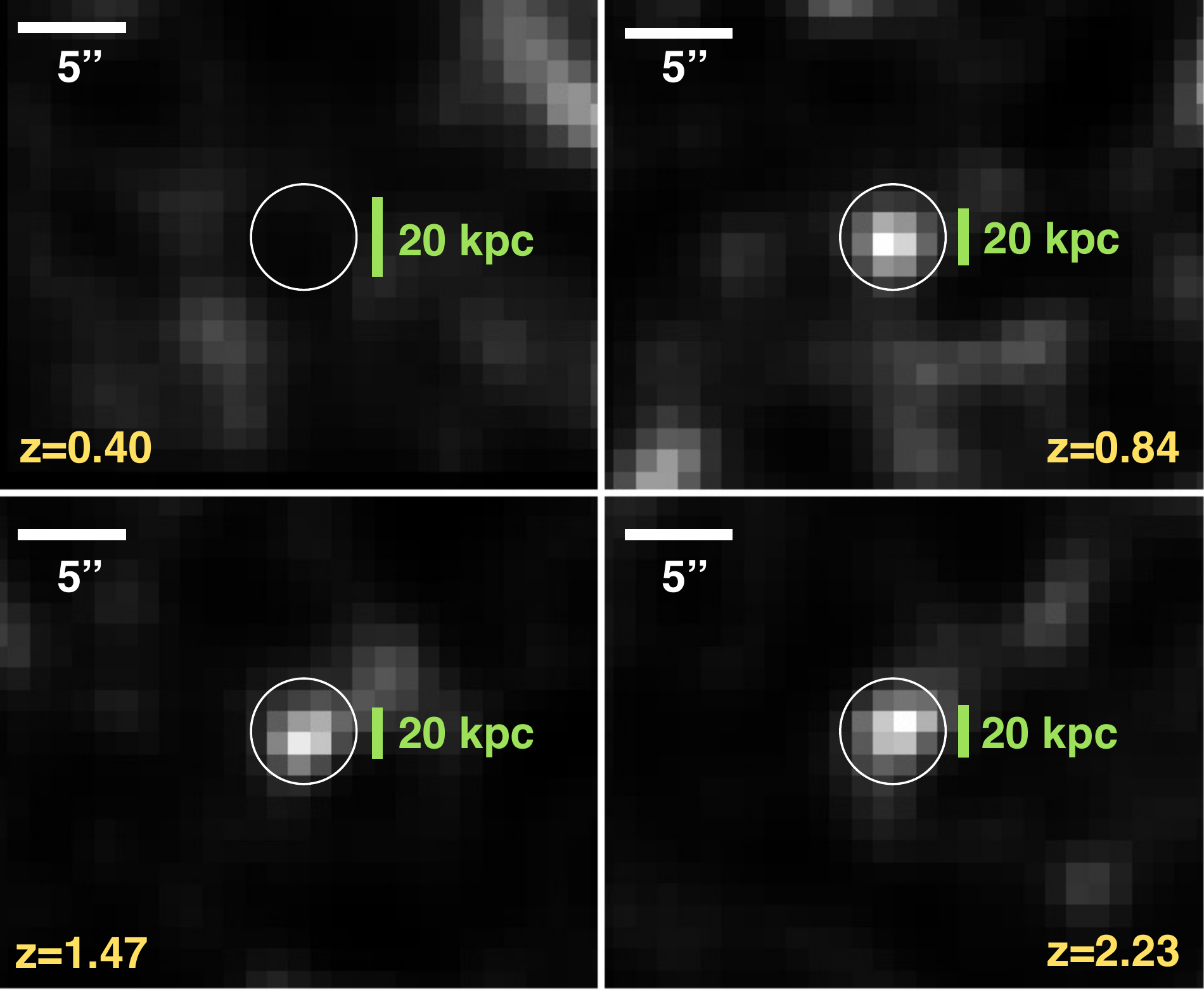}
	\caption{Stacking in the X-rays (\textit{Chandra}'s full band) for all our H$\alpha$ sources within the C-COSMOS coverage, in each our redshift slices. The results show high S/N detections at every redshift except for $z=0.4$. It is worth noting, however, that the sample at $z=0.4$ is much smaller and has much lower stellar mass and SFR on average than the other redshifts considered, and fails to encompass the rare luminous objects like AGN (see Figure \ref{fig:subfigures}), since it comes from a much smaller volume than the samples at higher redshifts. The images in this figure have been smoothed for easier inspection.}
	\label{fig:Stackoriginal}
\end{figure}

The vast majority of our H$\alpha$ emitters ($\sim98$\%) are undetected in the X-rays for the current C-COSMOS flux limit. This is expected given that the \textit{Chandra} sensitivity limit is $>10^{-16}$\,erg\,s$^{-1}$\,cm$^{-2}$. Thus, only relatively luminous AGN are expected to be X-ray detected, while our sample is strongly dominated by typical star-forming galaxies. However, we can rely on stacking in order to study the overall population of typical H$\alpha$ selected galaxies below the X-ray detection limit and recover much lower black hole accretion activity. In order to stack our samples of H$\alpha$ emitters, per redshift, we use the full energy band of C-COSMOS (0.5-7 keV) and start by cutting-out a square of $10''\times10''$ centred on each source. We adopt a stacking radius of 2$''$ (the area radius from which we extract the counts for the fluxes). These values were obtained by going through different values for the radius, selecting the ones that maximised the signal-to-noise (S/N) ratio (see \citealt{Lehmer2007} for details) and taking the mean. When stacking, we use all sources (both detected and non-detected), allowing us to include the entire population. \textit{Chandra}'s PSF changes with the distance to the pointings, causing deformation of sources. However, the effect of the changing PSF is minimal when compared with the error bars and uncertainties inherent to the FIR analysis. As such we did not apply a correction to this effect and instead estimated the background contribution by taking the standard deviation of the pixel counts in a randomized number of areas of the same size of the stacking area, making sure these would fall outside the vicinity of the stacking radius, in order to counter the possible presence of sources distorted by the changes in \textit{Chandra}'s PSF.

To convert background subtracted counts into fluxes we divided them by the mean exposure time multiplied by the conversion factor ($CR\times10^{-11}$ erg cm$^{-2}$ s$^{-1}$ (counts s$^{-1}$)$^{-1}$, where $CR$ is the count rate) assuming a power law of photon index $\Gamma=1.4$ and a Galactic absorption $N_H = 2.7\times10^{20}$ as in \cite{Elvis2009}. A photon index of 1.4 is appropriate for faint galaxies \citep[see][]{Alexander2003}, as we expect star-forming galaxies to be. Finally, all images were background subtracted. The estimation of the luminosities was done following: 

\begin{equation}
\rm L_X= 4 \pi {d_L}^2 f_X (1+z)^{\Gamma -2}\, (erg\,s^{-1}),
\end{equation}
where $\rm d_L$ is the luminosity distance, $\rm f_X$ is the flux in the X-ray band, $z$ is the redshift and $\Gamma$ is the photon index, assumed to be 1.4.

Figure \ref{fig:Stackoriginal} shows the results of the stacking for the four redshifts. There are clear detections for $z=0.84$, $z=1.47$ and $z=2.23$. For $z=0.4$ the S/N is much lower. This is not surprising, as i) this is the smallest sample and particularly because ii) the sources in the $z=0.4$ (due to the much smaller volume probed, see \S \ref{Hizels_survey}) are typically much lower luminosity and have lower stellar masses than those at higher redshift.

%
%
%
\begin{table*}
\centering
\caption[]{Quantities estimated for the stacked sources. Fluxes and luminosities in the X-ray band and estimated black hole accretion rates from these quantities were estimated from C-COSMOS. SFR estimated from the FIR luminosities as determined by \citet{Alasdair} and from radio data from VLA-COSMOS.}
\begin{tabular}{@{}cccccccccc@{}}
\hline
Source ID/Filter & $z$ & log Flux & log Luminosity & log Luminosity IR & SFR & SFR & $\rm \dot{M}_{\rm BH}$ & log [$\rm \dot{M}_{\rm BH}$ / SFR] (FIR)\\
    &  & (X-rays) & (X-rays) & (FIR) & (FIR)& (Radio) & & (X-rays)& \\
    &  & erg\,s$^{-1}$\,cm$^{-2}$  & erg\,s$^{-1}$ &  (L$_{\odot})$ & M$_{\odot}$\,yr$^{-1}$ &M$_{\odot}$\,yr$^{-1}$ & M$_{\odot}$\,yr$^{-1}$ & \\
\hline
NB921 & 0.4 & $<-15.4$ & $<41.25$ & 10.4$\pm$0.26 & 2$^{+1.6}_{-0.9}$ &1.5$^{+0.5}_{-0.2}$& $<0.0006$ & $<-3.55$\\  
NBJ & 0.85 & $-15.26\pm 0.12$ & $42.12\pm0.12$ & 11.1$\pm$0.23 &  13$^{+8.8}_{-5.2}$ & 10.5$^{+0.7}_{-0.6}$ & 0.004 $\pm$ 0.001 & -3.51$\pm$0.3\\  
NBH & 1.47 & $-15.06\pm 0.07$ & $42.83 \pm 0.07$ & 11.5$\pm$0.23 & 32$^{+21.7}_{-13.4}$ & 62$^{+3}_{-2.7}$& 0.02 $\pm$ 0.004 & -3.20$\pm$0.28\\   
NBK & 2.23 & $-15.33 \pm 0.12$ & $42.94\pm 0.12$ & 11.6$\pm$0.42& 40$^{+64.7}_{-24.9}$ & 21$^{+1.4}_{-1.3}$ & 0.03 $\pm$ 0.01 & -3.10$\pm$0.3\\  
\hline
\end{tabular}
\label{table:stacsourc}
\end{table*}

\subsubsection{Black hole accretion rate from X-ray luminosity}

We use the X-ray luminosity to estimate the rate at which the supermassive black hole at the centre of galaxies is accreting matter:
\begin{equation}
\dot{M}_{\rm BH}=\frac{(1-\epsilon)L^{AGN}_{bol}}{\epsilon c^2} (\rm{M_{\odot}\,yr^{-1}}),
\end{equation}
where $\dot{M}_{BH}$ is the accretion rate of the black hole, $\epsilon$ is the accretion efficiency, $L^{AGN}_{bol}$ is the bolometric luminosity of the AGN, obtained by multiplying the X-ray luminosity by 22.4 \citep{Lehmer2013, Vasudevan2007}, and $c$ is the speed of light. We find that our typical star-forming galaxies have accretion rates that rise with increasing redshift, from $\approx0.004$\,M$_{\odot}$\,yr$^{-1}$ at $z=0.84$ to $\approx0.03$\,M$_{\odot}$\,yr$^{-1}$ at $z=2.23$. When extracting the accretion rates from the X-ray luminosities, we estimated the correction that would have to be taken into account from the contribution to the X-ray emission by SF. This correction was estimated following \cite{Lehmer2016}:
\begin{equation}
\log{L_X} = A+B \log{(SFR)}+C \log{(1+z)}
\end{equation}
where A, B and C have the values $39.82\pm0.05$, $0.63\pm0.04$ and $1.31\pm0.11$ respectively. The correction turned out to be at most $\sim$0.05\% of the total BH accretion, much less than the uncertainties in quantities like SFR and actual BHAR and, as such, we do not take it into account. It also seems to evolve with galactic stellar mass, growing as the mass grows and following $L_X = 1.44(SFR)-0.45$ with $\chi^2=1.8$ when fitted to a linear relation throught the least-squares method. This evolution of the contribution to the X-rays from stars is not surprising, as the SFR also grows with stellar mass (see \ref{section:ratiomass} and \ref{fig:SFRBHARMass}).

%
%
%
%
\begin{table*}
 \centering
  \caption{Number of H$\alpha$ emitters classified as possible and likely AGN according to the selections mentioned in Section \ref{AGN_selection}.}
  \begin{tabular}{@{}cccccc@{}}
  \hline
  Method  & $z=0.4$ & $z=0.84$ & $z=1.47$ & $z=2.23$ & Total\\
    \hline
  X-ray Counterpart (C-COSMOS)  & 1 & 7 & 4 & 5 & 18 \\
  X-ray AGN Fraction & $3\pm2$\% & $3\pm2$\% & $3\pm2$\% & $2\pm1$\% & $3\pm2$\% \\ %
    \hline
  Radio Counterpart (VLA-COSMOS)  & 1 & 11 & 7 & 9 & 28 \\  
  \hline  
  Sources retained for stacking (X-rays) & 35 & 224 & 137 & 276 & 672 \\
  Sources retained for stacking (Radio)  & 35 & 214 & 132 & 268 & 649 \\
  Sources retained for stacking (FIR) & 35 & 224 & 136 & 276 & 671 \\
  \hline
\end{tabular}
\label{table:AGN}
\end{table*}

%
%
%
%
\begin{figure}
\centering
\includegraphics[width=8cm]{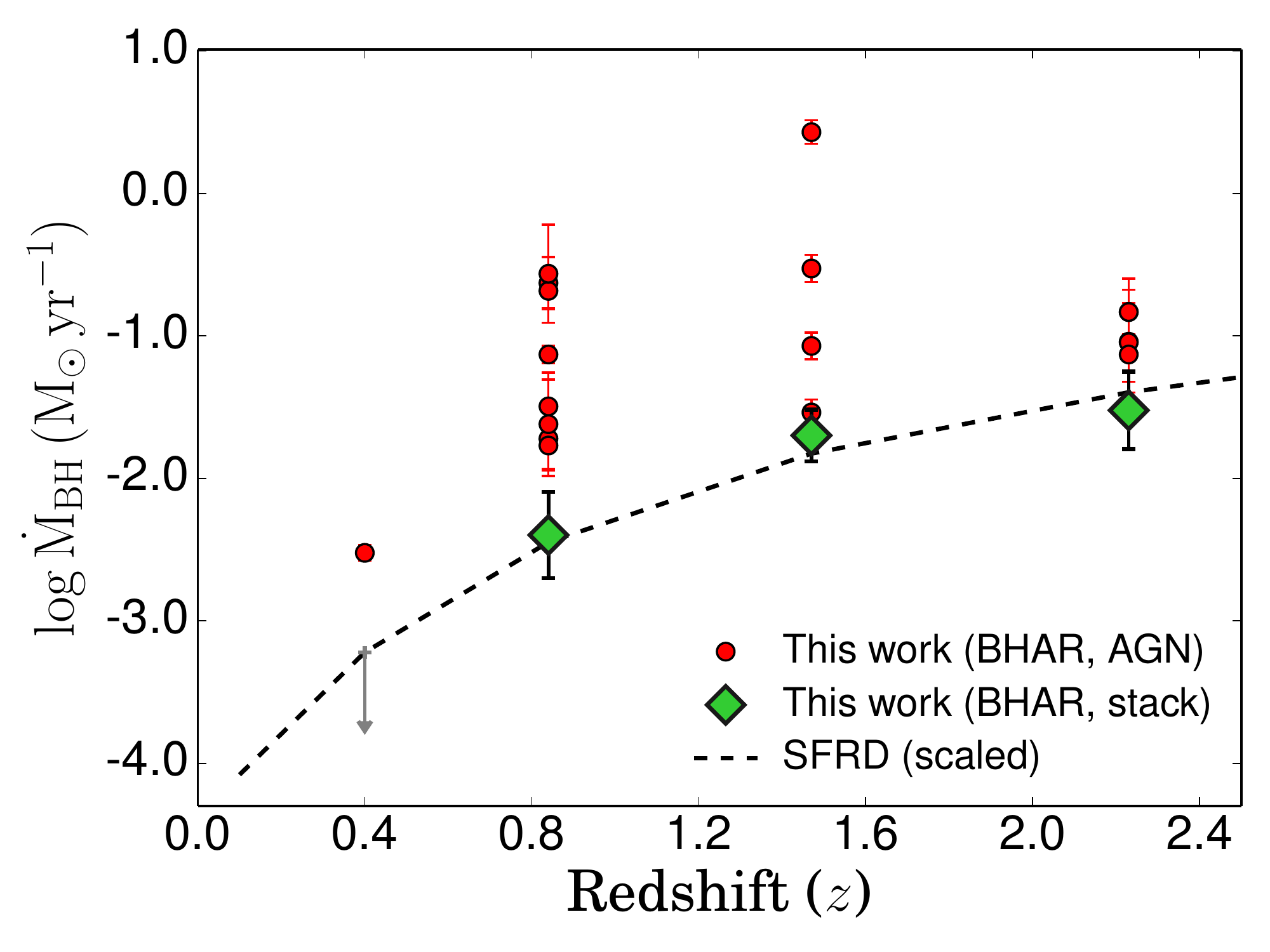}
\caption{The evolution of black hole accretion rates ($\rm \dot{M}_{\rm BH}$), for individually detected (in the X-rays) AGNs and for the stacks of the full samples. We compare those with a scaled evolution of the star formation rate density, SFRD \citep[][]{Sobral2013a}. The SFRD has been scaled to coincide with the $\rm \dot{M}_{\rm BH}$ at $z=0.4$. The results show that the $\rm \dot{M}_{\rm BH}$ grows with redshift, starting to plateau at z $\sim$ 2.23 and that the SFRD evolves in a very similar way to the accretion rate of the BHs, starting to stabilise at around the same redshifts. The grey down arrow represents a non-detection for the z=0.4 stack.}
\label{BHARz2015}
\end{figure}

\section{Results}

\subsection{The cosmic evolution of black hole accretion rates}

We find that $\rm \dot{M}_{\rm BH}$ rises with increasing redshift as shown in Figure \ref{BHARz2015}. However, from $z=1.47$ to $z=2.23$, even though the accretion rate still rises, it does so less steeply.  This is consistent with the results in the literature: \cite{Aird2010} finds the peak of AGN luminosity density to be at $z=1.2\pm0.1$. We compare this redshift evolution with the evolution of the star formation rate density, also shown in Figure \ref{BHARz2015}. We use the results from \cite{Sobral2013a, Sobral2014} and scale them arbitrarily to look for any potential differences and/or similarities between the evolution of SFRD and $\rm \dot{M}_{\rm BH}$ across cosmic time. Our scaling clearly reveals that star-forming galaxies form stars at a much higher rate than they grow their black holes ($\sim$ 3.3 orders of magnitude faster), but the relative evolution seems to be the same across redshift. We explore this further in Section \ref{ref_galx_black}. We also show the accretion rates computed for each individual X-ray AGN, which reveal large scatter (likely due to the high variability of AGN), but that generally agree with the trend of the global population.

\subsection{The dependence of $\rm \dot{M}_{\rm BH}$/SFR on stellar mass}\label{section:ratiomass}

\begin{figure}
\centering
\includegraphics[width=8cm]{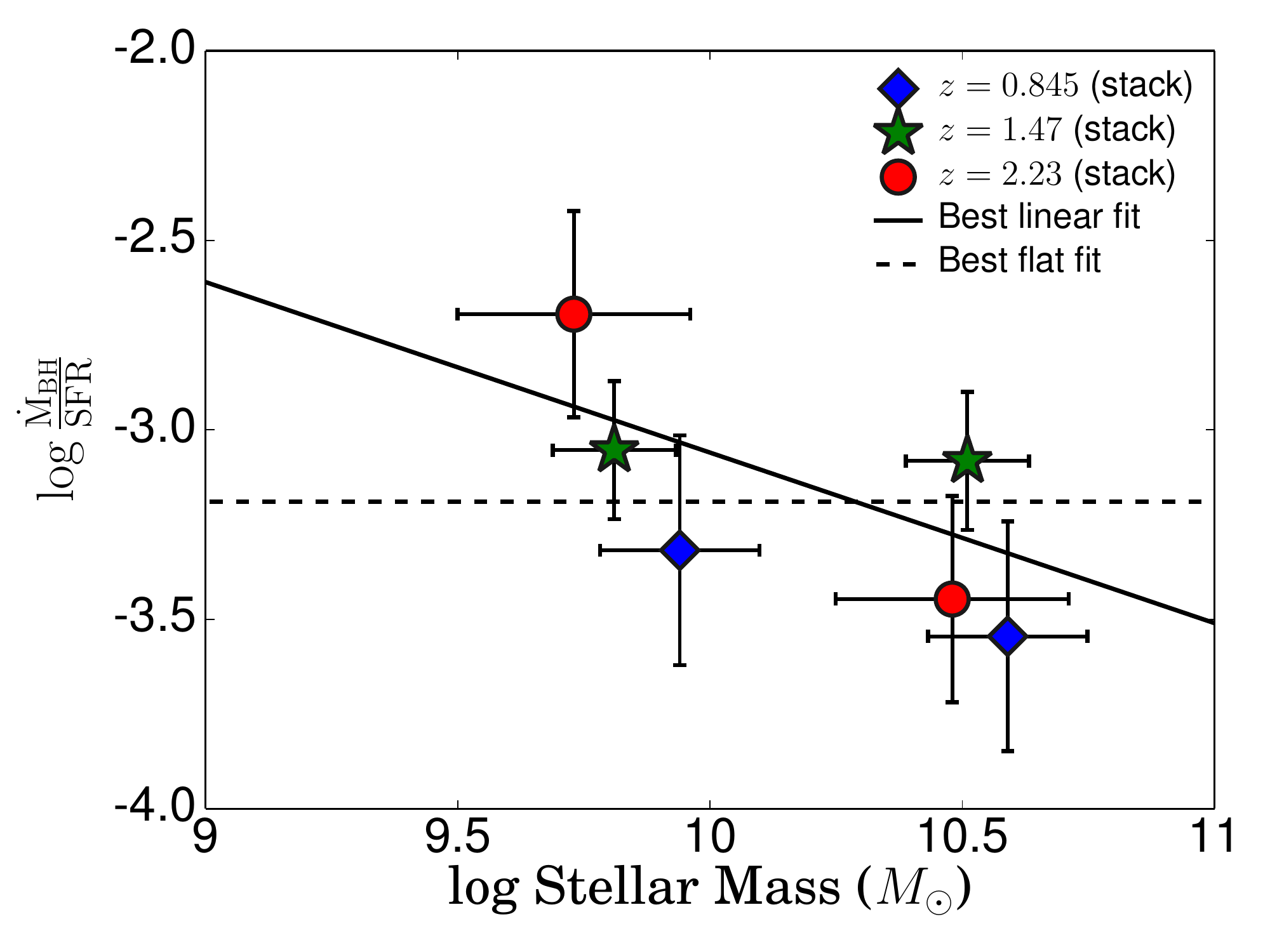}
\caption{The black hole accretion rate/SFR ratio ($\rm \dot{M}_{\rm BH}$/SFR) vs stellar mass for typical star-forming galaxies. The $\rm \dot{M}_{\rm BH}$/SFR ratio seems to generally decrease with stellar mass, indicating that more massive star-forming galaxies grow faster than their black holes compared to the least massive ones. The solid black line represents the best linear fit for ($\log(\dot{M_{\rm BH}}/SFR)=-0.45\log(M) + 1.44$; reduced $\chi ^2 = 1.8$). The dashed line represents the best fit for a flat relation (reduced $\chi ^2 = 2.8$).}\label{MoverSFRMass2015}
\end{figure}

Using the results from the FIR analysis we are able to estimate SFRs which should be independent of AGN activity. We use those to determine the ratio between the black hole accretion rate and SFR ($\rm \dot{M}_{\rm BH}$/SFR). Figure \ref{MoverSFRMass2015} shows how $\rm \dot{M}_{\rm BH}$/SFR depends on stellar mass \citep[stellar masses computed in][]{Sobral2014} for the three different redshifts where we can easily split our samples. We find that a linear relation with a slope of $-0.45$ provides the best fit (see Figure \ref{MoverSFRMass2015}). We find that both $\rm \dot{M}_{\rm BH}$ and SFR increase with stellar mass, but SFR seems to rise slightly faster with stellar mass than $\rm \dot{M}_{\rm BH}$ (see \ref{fig:SFRBHARMass}). However, our results are still fully consistent with a completely flat relation (only $\sim1$\,$\sigma$ away from a flat relation). This may be a sign that the BH accretion and SF of our typical star-forming galaxies evolve at equivalent rates across cosmic time, as we do not find any strong evidence for evolution with cosmic time either. Given that the peak of BH and SF activity is thought to occur at redshifts between $z\sim1-2$, this constancy seems to support the idea that the central supermassive BHs and SF mechanism form a single way of regulating galaxy growth, as opposed to one mechanism taking over the other at set intervals in time. It should be noted, however, that other works, such as \cite{Kormendy2013} and \cite{Rodighiero2015}, have found a different evolution of the ratio with stellar mass with the ratio increasing with the stellar mass, with \cite{Rodighiero2015} finding that the ratio between the X-ray luminosity and SFR scales as $\log{(L_X/SFR)} \propto {M_*}^{0.43 \pm 0.09}$.

%
%
%
%

\subsection{Relative black hole-galaxy growth and its redshift evolution} \label{ref_galx_black}

\begin{figure}
\centering
\includegraphics[width=8cm]{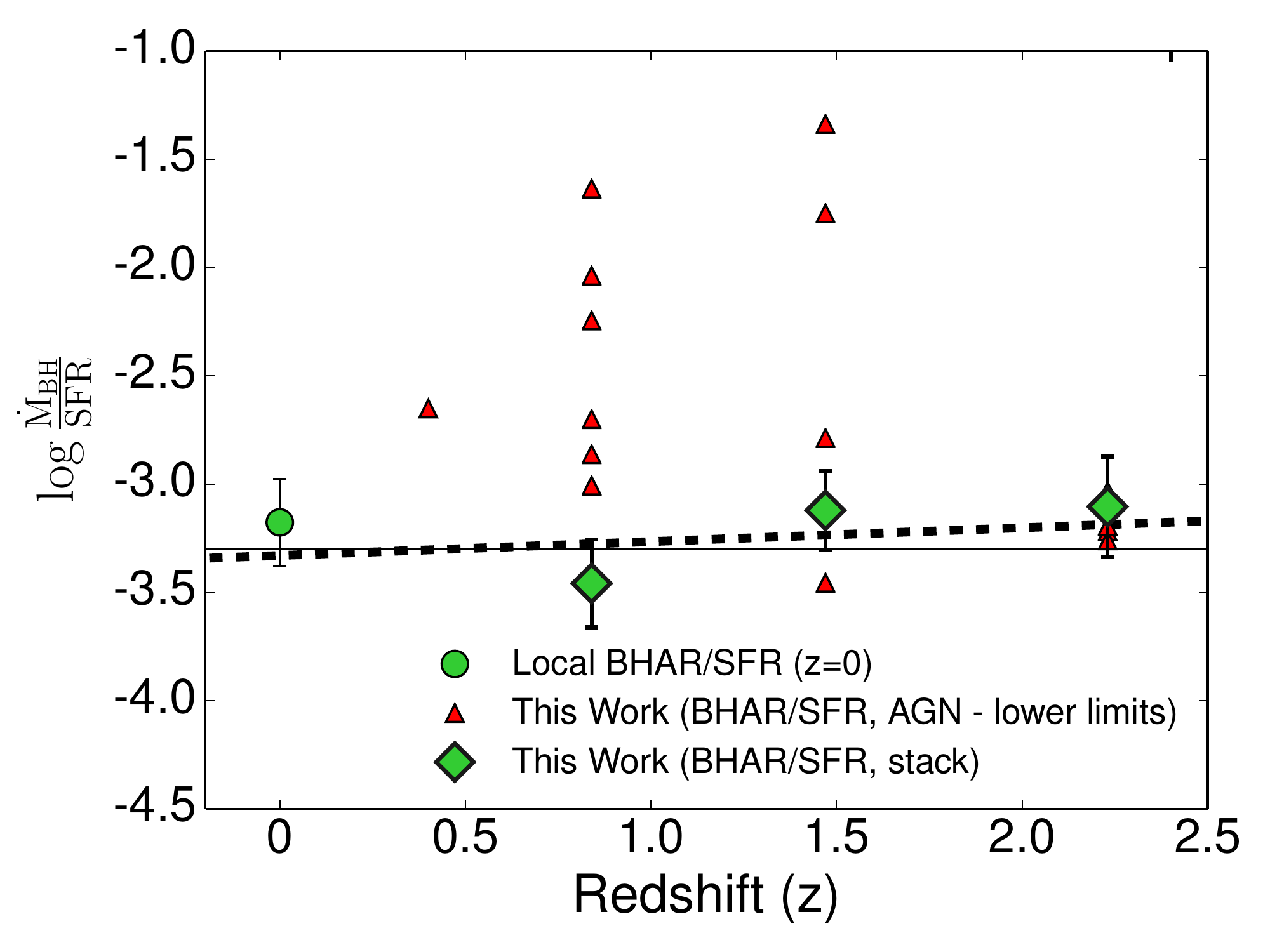}
\caption{The evolution of the black hole accretion rate/SFR ratio ($\rm \dot{M}_{\rm BH}$/SFR) from $z=0$ to $z=2.23$. Our results show little to no evolution in $\rm \dot{M}_{\rm BH}$/SFR over the last 11\,Gyrs of cosmic time. The grey line represents a constant relation, while the dashed line is the best fit (less than 1\,$\sigma$ away from a flat relation). The ratio for the stacking remains approximately the same for all redshifts ($-3.3\pm0.2$), being consistent with the measured $\rm \dot{M}_{\rm BH}$/SFR value for the local Universe. This seems to shows that typical star-forming galaxies form stars much faster than their BHs grow, with such difference being approximately constant across cosmic time. We also show lower limits for individual sources detected in the X-rays there; these show a large scatter with a potential peak at $z\sim1-1.5$.}\label{MoverSFRredshift}
\end{figure}

Figure \ref{MoverSFRredshift} shows how the ratio between the black hole accretion rate and SFR evolves across cosmic time (see also Table \ref{table:stacsourc}). We find that the ratio between black hole and galaxy growth is very low and is surprisingly constant across redshift, $\sim10^{-3.3}$. We thus find little to no evolution from $z=2.23$ to $z=0$. We investigate a potential linear fit and compare it to a flat relation (no evolution in redshift). Our results prefer a slope that is completely consistent, within less than 1\,$\sigma$ with a flat relation (see Figure \ref{MoverSFRredshift}). This is consistent with previous results: \cite{Mullaney2012} find a flat, non-evolving relation between SFR and $\rm \dot{M}_{\rm BH}$, also maintaining a ratio of $\sim10^{-3}$ for redshifts of $0.5<z<2.5$. This was interpreted as a sign that the SFR and $\rm \dot{M}_{\rm BH}$ evolve equivalently throughout cosmic history, in tight relation with one another and with practically no ``lag" between the two, a conclusion supported by \cite{Chen2013}, who found an almost linear correlation between the $\rm \dot{M}_{\rm BH}$ and SFR of star forming galaxies for redshifts $0.25<z<0.8$.

We can only provide lower limits for the X-ray AGN, but those provide evidence for strong scatter, likely driven by strong AGN variability. Such scatter/variability may well be higher at $z\sim1-2$ than at lower redshifts. Not only is the BH more active in the X-ray AGN, with accretion rates at least an order of magnitude higher than the stacked sources (compare Tables \ref{table:detsourc} and \ref{table:stacsourc}), but the AGN activity itself may be having an effect on the SFR. We note that our results are consistent with those presented by \cite{Lehmer2013}. The stacked sources show an accretion rate/SFR ratio typical of star forming galaxies, while the directly detected sources present a ratio in line with AGN (Figure \ref{MoverSFRredshift}). This is expected: throughout their lives, galaxies are thought to move above or below the local ratio depending on their AGN activity and SFR.

We note that our results do not depend on the choice of SFR indicator. Particularly, the SFRs obtained from e.g. the radio are in line with those determined with infrared luminosity ($\sim$1 M$_{\odot}$\,yr$^{-1}$ for $z=0.4$ and $\sim$20 M$_{\odot}$\,yr$^{-1}$ for $z=2.23$), and are also similar to those derived from H$\alpha$. However, we use FIR SFRs because they should be less affected by AGN activity than the radio (and H$\alpha$). Even though we excluded radio sources more luminous than 10$^{22}$\,W\,Hz$^{-1}$ (when obtaining radio SFRs), we may still get some AGN contamination. Furthermore, even though SF-related radio emission has its origins in the supernovae of massive stars (whose life-times are comparable to the duration of the star formation period), the electrons responsible for the radiation continue emitting for periods of time that reach up to $\sim$100 million years after the original stars exploded. While this ``persistence" of emission depends on factors like the density of the surrounding environment, it means that SFRs from the radio trace timescales that are longer than those from FIR and H$\alpha$.

%
%
%
%

\section{Conclusion}

We have investigated the relative growth of H$\alpha$-selected star-forming galaxies and their supermassive black holes across a redshift range of $0.4\leq z \leq 2.23$ by making use of the HiZELS sample and the wealth of data available for the COSMOS field. We determined the black hole accretion rate of galaxies from their X-ray luminosities and their SFR from their luminosity in the far-infrared. In this manner, we were able to estimate the $\rm \dot{M}_{\rm BH}$/SFR ratio for typical star-forming galaxies and how that evolves with cosmic time.

Only $\sim3$\% of the H$\alpha$-selected star-forming population are detected in the X-rays as AGN. Our results are in line with the results from the literature: \cite{Garn2010} found that only a few per cent of the H$\alpha$ emitters at $z=0.84$ are detected in the X-rays. \cite{Sobral2016} found similar results, with X-ray-detected AGN fractions that varied from 1\% to 2-3\% for redshifts $0.8 \leq z\leq 2.23$. Our X-ray AGN fractions are 3\% for the redshifts $z=0.4-1.47$ and 2\% for $z=2.23$. This implies that there is no significant evolution of the X-ray AGN fraction with redshift. Our results also complement those from \cite{Sobral2016}, who estimated AGN fractions at $z=0.84-2.23$ for the most luminous H$\alpha$ emitters and found little to no evolution with redshift.

The FIR SFRs in our sample range from $\sim2$\,M$_{\odot}$\,yr$^{-1}$ to $\sim40$\,M\,$_{\odot}$\,yr$^{-1}$, from $z=0.4$ to $z=2.23$ \citep[][]{Alasdair}. This is in good agreement with the H$\alpha$ SFRs \citep[see e.g.][]{Swinbank2012,Sobral2014}. The $\rm \dot{M}_{\rm BH}$ we obtain are generally a thousandth of the SFRs of the galaxies we studied, in line with results from \cite{Lehmer2013} for star-forming galaxies at $z=2.23$. The black hole accretion rates rise with redshift from $\rm \dot{M}_{BH}\sim0.004$\,M$_{\odot}$\,yr$^{-1}$ at $z=0.8$ to $\rm \dot{M}_{BH}\sim0.03$\,M$_{\odot}$\,yr$^{-1}$ at $z=2.23$. The rising of the $\rm \dot{M}_{\rm BH}$ may be steeper until $z=1.47$. Interestingly, the SFRD evolves in a very similar way to the $\rm \dot{M}_{\rm BH}$, starting to stabilise at around the same redshifts: the $\rm \dot{M}_{\rm BH}$ evolution starts to ``flatten" at $1.47<z<2.23$ \citep[e.g.][]{Sobral2013a}, something that is supported in the literature, as \cite{Aird2010} has found that the peak of X-ray luminosity density is located at $z = 1.2 \pm 0.1$.

Our $\rm \dot{M}_{\rm BH}$/SFR ratio is observed to have little to no evolution with redshift, being approximately $\sim10^{-3.3}$ between $z=0$ and $z=2.23$. This little to no evolution across redshift suggests that $\rm \dot{M}_{\rm BH}$ and SFRs of our typical star-forming galaxies evolve at similar rates across cosmic time. Our results are thus in good agreement with the ones in the literature. Several authors have noted that the $\rm \dot{M}_{\rm BH}$ and SFR ratio has been independent of cosmic time for the last $\sim10$\,Gyrs, with a value of $\sim10^{-3.2}$ \citep[see e.g.][]{Shankar2009, HopkinsBeacom2006, HeckmanBest2014}. It is worth noting that, although our results favor a scenario where the black holes and their host galaxies grow simultaneously as a whole, they do not imply that this is necessarily the case on a galaxy by galaxy basis. Nevertheless, the little to no evolution of $\rm \dot{M}_{\rm BH}$/SFR across cosmic time suggests that the processes that fuel $\rm \dot{M}_{\rm BH}$ and SFR have remained the essentially the same (or correlated) over cosmic time \citep[see, e.g.][]{Heckman2004, Mullaney2012}. However, understanding and explaining these physical processes in detail (feedback, gas stability and availability) is still a very important open question.

We also find that $\rm \dot{M}_{\rm BH}$/SFR may decline slightly with increasing stellar mass, although very weakly. This specific relation is interesting because the canonical interpretation of the influence of AGN and star formation in galaxy evolution is that AGN generally dominate in more massive galaxies whereas in less massive galaxies star formation starts playing a more important role. The fact that $\rm \dot{M}_{\rm BH}$/SFR depends so little on galaxy mass could indicate that BH activity and SFR form a combined mechanism for the regulation of galaxy growth, as opposed to simply one mechanism taking over the other at set intervals in time, but this is currently very uncertain.

As for the directly detected sources in the X-rays (X-ray AGN), they show very significant scatter. They seem to deviate from the behaviour of the full population, revealing $\rm \dot{M}_{\rm BH}$/SFR ratios of $>10^{-3.5}$ to $>10^{-1.2}$. This is not a surprising result, since AGN activity is highly variable and the BH growth may exceed SFR and vice-versa on short timescales \citep[e.g.][]{Alexander2008, Targett2012}.

Future work would need to focus on extending this study to other surveys as well as trying to understand how SF and BH activity might constrain the evolution of the galaxies they happen in. The further use of ALMA to probe gas outflows in AGN and SF galaxies would allow us to get a much more detailed idea of whether these processes affect galaxies differently and let us better understand how AGN and SF influence galaxy growth and themselves.

\section*{Acknowledgements}

We thank the reviewer for their helpful comments and suggestions.
JC and DS acknowledge financial support from the Netherlands Organisation for Scientific research (NWO) through a Veni fellowship, from FCT through a FCT Investigator Starting Grant and Start-up Grant (IF/01154/2012/CP0189/CT0010) and from FCT grant PEst-OE/FIS/UI2751/2014. JC also acknowledges a Lancaster University PhD studentship. PNB is grateful for support from the UK STFC via grant ST/M001229/1. CMH, APT and IRS acknowledge support from STFC(ST/L00075X/1). APT and IRS also acknowledge support from the ERC advanced Grant DUSTYGAL (321334). In addition, IRS acknowledges support from a Royal Society/Wolfson Merit Award. This research has made use of NASA's Astrophysics Data System.

\bibliographystyle{mnras}
\bibliography{BHGalaxyGrowth}

\begin{thebibliography}{}
\makeatletter
\relax
\def\mn@urlcharsother{\let\do\@makeother \do\$\do\&\do\#\do\^\do\_\do\%\do\~}
\def\mn@doi{\begingroup\mn@urlcharsother \@ifnextchar [ {\mn@doi@}
  {\mn@doi@[]}}
\def\mn@doi@[#1]#2{\def\@tempa{#1}\ifx\@tempa\@empty \href
  {http://dx.doi.org/#2} {doi:#2}\else \href {http://dx.doi.org/#2} {#1}\fi
  \endgroup}
\def\mn@eprint#1#2{\mn@eprint@#1:#2::\@nil}
\def\mn@eprint@arXiv#1{\href {http://arxiv.org/abs/#1} {{\tt arXiv:#1}}}
\def\mn@eprint@dblp#1{\href {http://dblp.uni-trier.de/rec/bibtex/#1.xml}
  {dblp:#1}}
\def\mn@eprint@#1:#2:#3:#4\@nil{\def\@tempa {#1}\def\@tempb {#2}\def\@tempc
  {#3}\ifx \@tempc \@empty \let \@tempc \@tempb \let \@tempb \@tempa \fi \ifx
  \@tempb \@empty \def\@tempb {arXiv}\fi \@ifundefined
  {mn@eprint@\@tempb}{\@tempb:\@tempc}{\expandafter \expandafter \csname
  mn@eprint@\@tempb\endcsname \expandafter{\@tempc}}}

\bibitem[\protect\citeauthoryear{{Aird}, {Nandra}  et~al.}{{Aird}
  et~al.}{2010}]{Aird2010}
{Aird} J.,  {Nandra} K.,   et~al., 2010, \mn@doi [\mnras]
  {10.1111/j.1365-2966.2009.15829.x}, \href
  {http://adsabs.harvard.edu/abs/2010MNRAS.401.2531A} {401, 2531}

\bibitem[\protect\citeauthoryear{{Alexander} et~al.,}{{Alexander}
  et~al.}{2003}]{Alexander2003}
{Alexander} D.~M.,  et~al., 2003, \mn@doi [AJ] {10.1086/346088}, \href
  {http://adsabs.harvard.edu/abs/2003AJ....125..383A} {125, 383}

\bibitem[\protect\citeauthoryear{{Alexander} et~al.,}{{Alexander}
  et~al.}{2008}]{Alexander2008}
{Alexander} D.~M.,  et~al., 2008, \mn@doi [\aj] {10.1088/0004-6256/135/5/1968},
  \href {http://adsabs.harvard.edu/abs/2008AJ....135.1968A} {135, 1968}

\bibitem[\protect\citeauthoryear{{Best}, {Kauffmann}, {Heckman}, {Brinchmann},
  {Charlot}, {Ivezi{\'c}}  \& {White}}{{Best} et~al.}{2005}]{Best2005}
{Best} P.~N.,  {Kauffmann} G.,  {Heckman} T.~M.,  {Brinchmann} J.,  {Charlot}
  S.,  {Ivezi{\'c}} {\v Z}.,   {White} S.~D.~M.,  2005, \mn@doi [MNRAS]
  {10.1111/j.1365-2966.2005.09192.x}, \href
  {http://adsabs.harvard.edu/abs/2005MNRAS.362...25B} {362, 25}

\bibitem[\protect\citeauthoryear{{Best}, {Kaiser}, {Heckman}  \&
  {Kauffmann}}{{Best} et~al.}{2006}]{Best2006}
{Best} P.~N.,  {Kaiser} C.~R.,  {Heckman} T.~M.,   {Kauffmann} G.,  2006,
  \mn@doi [MNRAS] {10.1111/j.1745-3933.2006.00159.x}, \href
  {http://adsabs.harvard.edu/abs/2006MNRAS.368L..67B} {368, L67}

\bibitem[\protect\citeauthoryear{{Best} et~al.,}{{Best}
  et~al.}{2013}]{Best2010}
{Best} P.,  et~al., 2013, \mn@doi [ASSP] {10.1007/978-94-007-7432-2_22}, \href
  {http://adsabs.harvard.edu/abs/2013ASSP...37..235B} {37, 235}

\bibitem[\protect\citeauthoryear{{Bondi}, {Ciliegi}, {Schinnerer}, {Smol{\v
  c}i{\'c}}, {Jahnke}, {Carilli}  \& {Zamorani}}{{Bondi}
  et~al.}{2008}]{Bondi2008}
{Bondi} M.,  {Ciliegi} P.,  {Schinnerer} E.,  {Smol{\v c}i{\'c}} V.,  {Jahnke}
  K.,  {Carilli} C.,   {Zamorani} G.,  2008, \mn@doi [ApJ] {10.1086/589324},
  \href {http://adsabs.harvard.edu/abs/2008ApJ...681.1129B} {681, 1129}

\bibitem[\protect\citeauthoryear{{Bower}, {Benson}, {Malbon}, {Helly}, {Frenk},
  {Baugh}, {Cole}  \& {Lacey}}{{Bower} et~al.}{2006}]{Bower2006}
{Bower} R.~G.,  {Benson} A.~J.,  {Malbon} R.,  {Helly} J.~C.,  {Frenk} C.~S.,
  {Baugh} C.~M.,  {Cole} S.,   {Lacey} C.~G.,  2006, \mn@doi [MNRAS]
  {10.1111/j.1365-2966.2006.10519.x}, \href
  {http://adsabs.harvard.edu/abs/2006MNRAS.370..645B} {370, 645}

\bibitem[\protect\citeauthoryear{{Brandt} \& {Alexander}}{{Brandt} \&
  {Alexander}}{2015}]{Brandt2015}
{Brandt} W.~N.,  {Alexander} D.~M.,  2015, \mn@doi [\aapr]
  {10.1007/s00159-014-0081-z}, \href
  {http://adsabs.harvard.edu/abs/2015A%26ARv..23....1B} {23, 1}

\bibitem[\protect\citeauthoryear{{Casali}, {Adamson}, {Alves de Oliveira}
  et~al.}{{Casali} et~al.}{2007}]{Casali2007}
{Casali} M.,  {Adamson} A.,  {Alves de Oliveira} C.,   et~al., 2007, \mn@doi
  [A\&A] {10.1051/0004-6361:20066514}, \href
  {http://adsabs.harvard.edu/abs/2007A26A...467..777C} {467, 777}

\bibitem[\protect\citeauthoryear{{Chabrier}}{{Chabrier}}{2003}]{Chabrier2003}
{Chabrier} G.,  2003, \mn@doi [\pasp] {10.1086/376392}, \href
  {http://adsabs.harvard.edu/abs/2003PASP..115..763C} {115, 763}

\bibitem[\protect\citeauthoryear{{Chen} et~al.,}{{Chen}
  et~al.}{2013}]{Chen2013}
{Chen} C.-T.~J.,  et~al., 2013, \mn@doi [\apj] {10.1088/0004-637X/773/1/3},
  \href {http://adsabs.harvard.edu/abs/2013ApJ...773....3C} {773, 3}

\bibitem[\protect\citeauthoryear{{Davies}, {Maciejewski}, {Hicks}, {Tacconi},
  {Genzel}  \& {Engel}}{{Davies} et~al.}{2009}]{Davies2009}
{Davies} R.~I.,  {Maciejewski} W.,  {Hicks} E.~K.~S.,  {Tacconi} L.~J.,
  {Genzel} R.,   {Engel} H.,  2009, \mn@doi [ApJ]
  {10.1088/0004-637X/702/1/114}, \href
  {http://adsabs.harvard.edu/abs/2009ApJ...702..114D} {702, 114}

\bibitem[\protect\citeauthoryear{{Delvecchio}, {Gruppioni}
  et~al.}{{Delvecchio} et~al.}{2014}]{Delvecchio2014}
{Delvecchio} I.,  {Gruppioni} C.,   et~al., 2014, \mn@doi [\mnras]
  {10.1093/mnras/stu130}, \href
  {http://adsabs.harvard.edu/abs/2014MNRAS.439.2736D} {439, 2736}

\bibitem[\protect\citeauthoryear{{Delvecchio}, {Lutz}, {Berta}
  et~al.}{{Delvecchio} et~al.}{2015}]{Delvecchio2015}
{Delvecchio} I.,  {Lutz} D.,  {Berta} S.,   et~al., 2015, \mn@doi [MNRAS]
  {10.1093/mnras/stv213}, \href
  {http://adsabs.harvard.edu/abs/2015MNRAS.449..373D} {449, 373}

\bibitem[\protect\citeauthoryear{{Dunn} et~al.,}{{Dunn}
  et~al.}{2010}]{Dunn2010}
{Dunn} J.~P.,  et~al., 2010, \mn@doi [ApJ] {10.1088/0004-637X/709/2/611}, \href
  {http://adsabs.harvard.edu/abs/2010ApJ...709..611D} {709, 611}

\bibitem[\protect\citeauthoryear{{Elvis}, {Civano}, {Vignali}  et~al.}{{Elvis}
  et~al.}{2009}]{Elvis2009}
{Elvis} M.,  {Civano} F.,  {Vignali} C.,   et~al., 2009, \mn@doi [ApJS]
  {10.1088/0067-0049/184/1/158}, \href
  {http://adsabs.harvard.edu/abs/2009ApJS..184..158E} {184, 158}

\bibitem[\protect\citeauthoryear{{Ganguly} \& {Brotherton}}{{Ganguly} \&
  {Brotherton}}{2008}]{GangulyBrotherton}
{Ganguly} R.,  {Brotherton} M.~S.,  2008, \mn@doi [ApJ] {10.1086/524106}, \href
  {http://adsabs.harvard.edu/abs/2008ApJ...672..102G} {672, 102}

\bibitem[\protect\citeauthoryear{{Garn} et~al.,}{{Garn}
  et~al.}{2010}]{Garn2010}
{Garn} T.,  et~al., 2010, \mn@doi [MNRAS] {10.1111/j.1365-2966.2009.16042.x},
  \href {http://adsabs.harvard.edu/abs/2010MNRAS.402.2017G} {402, 2017}

\bibitem[\protect\citeauthoryear{{Geach}, {Smail}, {Best}, {Kurk}, {Casali},
  {Ivison}  \& {Coppin}}{{Geach} et~al.}{2008}]{Geach2008}
{Geach} J.~E.,  {Smail} I.,  {Best} P.~N.,  {Kurk} J.,  {Casali} M.,  {Ivison}
  R.~J.,   {Coppin} K.,  2008, \mn@doi [MNRAS]
  {10.1111/j.1365-2966.2008.13481.x}, \href
  {http://adsabs.harvard.edu/abs/2008MNRAS.388.1473G} {388, 1473}

\bibitem[\protect\citeauthoryear{{Geach} et~al.,}{{Geach}
  et~al.}{2013}]{Geach2013}
{Geach} J.~E.,  et~al., 2013, \mn@doi [\mnras] {10.1093/mnras/stt352}, \href
  {http://adsabs.harvard.edu/abs/2013MNRAS.432...53G} {432, 53}

\bibitem[\protect\citeauthoryear{{Geach} et~al.,}{{Geach}
  et~al.}{2014}]{Geach2014}
{Geach} J.~E.,  et~al., 2014, \mn@doi [Nature] {10.1038/nature14012}, \href
  {http://adsabs.harvard.edu/abs/2014Natur.516...68G} {516, 68}

\bibitem[\protect\citeauthoryear{{Geach} et~al.}{{Geach}
  et~al.}{2016}]{Geach2016}
{Geach} J. E.,  et~al., 2016, submitted, \href
  {http://adsabs.harvard.edu/abs/2016MNRAS.436.1130S} {}

\bibitem[\protect\citeauthoryear{{Genel} et~al.,}{{Genel}
  et~al.}{2014}]{Illustris2015}
{Genel} S.,  et~al., 2014, \mn@doi [\mnras] {10.1093/mnras/stu1654}, \href
  {http://adsabs.harvard.edu/abs/2014MNRAS.445..175G} {445, 175}

\bibitem[\protect\citeauthoryear{{Griffin} et~al.,}{{Griffin}
  et~al.}{2010}]{Griffin2010}
{Griffin} M.~J.,  et~al., 2010, \mn@doi [\aap] {10.1051/0004-6361/201014519},
  \href {http://adsabs.harvard.edu/abs/2010A%26A...518L...3G} {518, L3}

\bibitem[\protect\citeauthoryear{{Harrison} et~al.,}{{Harrison}
  et~al.}{2012}]{Harrison2012}
{Harrison} C.~M.,  et~al., 2012, \mn@doi [\apjl] {10.1088/2041-8205/760/1/L15},
  \href {http://adsabs.harvard.edu/abs/2012ApJ...760L..15H} {760, L15}

\bibitem[\protect\citeauthoryear{{Heckman} \& {Best}}{{Heckman} \&
  {Best}}{2014}]{HeckmanBest2014}
{Heckman} T.~M.,  {Best} P.~N.,  2014, \mn@doi [\araa]
  {10.1146/annurev-astro-081913-035722}, \href
  {http://adsabs.harvard.edu/abs/2014ARA%26A..52..589H} {52, 589}

\bibitem[\protect\citeauthoryear{{Heckman}, {Kauffmann}, {Brinchmann},
  {Charlot}, {Tremonti}  \& {White}}{{Heckman} et~al.}{2004}]{Heckman2004}
{Heckman} T.~M.,  {Kauffmann} G.,  {Brinchmann} J.,  {Charlot} S.,  {Tremonti}
  C.,   {White} S.~D.~M.,  2004, \mn@doi [\apj] {10.1086/422872}, \href
  {http://adsabs.harvard.edu/abs/2004ApJ...613..109H} {613, 109}

\bibitem[\protect\citeauthoryear{{Hopkins} \& {Beacom}}{{Hopkins} \&
  {Beacom}}{2006}]{HopkinsBeacom2006}
{Hopkins} A.~M.,  {Beacom} J.~F.,  2006, \mn@doi [\apj] {10.1086/506610}, \href
  {http://adsabs.harvard.edu/abs/2006ApJ...651..142H} {651, 142}

\bibitem[\protect\citeauthoryear{{Ibar} et~al.,}{{Ibar}
  et~al.}{2013}]{Ibar2013}
{Ibar} E.,  et~al., 2013, \mn@doi [\mnras] {10.1093/mnras/stt1258}, \href
  {http://adsabs.harvard.edu/abs/2013MNRAS.434.3218I} {434, 3218}

\bibitem[\protect\citeauthoryear{{Karim} et~al.}{{Karim}
  et~al.}{2011}]{Karim2011}
{Karim} A.,  et~al., 2011, \mn@doi [ApJ] {10.1088/0004-637X/730/2/61}, \href
  {http://adsabs.harvard.edu/abs/2011ApJ...730...61K} {730, 61}

\bibitem[\protect\citeauthoryear{{Kormendy} \& {Ho}}{{Kormendy} \&
  {Ho}}{2013}]{Kormendy2013}
{Kormendy} J.,  {Ho} L.~C.,  2013, \mn@doi [\araa]
  {10.1146/annurev-astro-082708-101811}, \href
  {http://adsabs.harvard.edu/abs/2013ARA%26A..51..511K} {51, 511}

\bibitem[\protect\citeauthoryear{{Lehmer} et~al.,}{{Lehmer}
  et~al.}{2007}]{Lehmer2007}
{Lehmer} B.~D.,  et~al., 2007, \mn@doi [\apj] {10.1086/511297}, \href
  {http://adsabs.harvard.edu/abs/2007ApJ...657..681L} {657, 681}

\bibitem[\protect\citeauthoryear{{Lehmer} et~al.,}{{Lehmer}
  et~al.}{2013}]{Lehmer2013}
{Lehmer} B.~D.,  et~al., 2013, \mn@doi [ApJ] {10.1088/0004-637X/765/2/87},
  \href {http://adsabs.harvard.edu/abs/2013ApJ...765...87L} {765, 87}

\bibitem[\protect\citeauthoryear{{Lehmer} et~al.,}{{Lehmer}
  et~al.}{2016}]{Lehmer2016}
{Lehmer} B.~D.,  et~al., 2016, \mn@doi [\apj] {10.3847/0004-637X/825/1/7},
  \href {http://adsabs.harvard.edu/abs/2016ApJ...825....7L} {825, 7}

\bibitem[\protect\citeauthoryear{{Lilly}, {Le Fevre}, {Hammer}  \&
  {Crampton}}{{Lilly} et~al.}{1996}]{Lilly96}
{Lilly} S.~J.,  {Le Fevre} O.,  {Hammer} F.,   {Crampton} D.,  1996, \mn@doi
  [ApJL] {10.1086/309975}, \href
  {http://adsabs.harvard.edu/abs/1996ApJ...460L...1L} {460, L1}

\bibitem[\protect\citeauthoryear{{Lutz} et~al.,}{{Lutz}
  et~al.}{2011}]{Lulz2011}
{Lutz} D.,  et~al., 2011, \mn@doi [\aap] {10.1051/0004-6361/201117107}, \href
  {http://adsabs.harvard.edu/abs/2011A%26A...532A..90L} {532, A90}

\bibitem[\protect\citeauthoryear{{Madau} \& {Dickinson}}{{Madau} \&
  {Dickinson}}{2014}]{Madau2014}
{Madau} P.,  {Dickinson} M.,  2014, \mn@doi [ARA\&A]
  {10.1146/annurev-astro-081811-125615}, \href
  {http://adsabs.harvard.edu/abs/2014ARA%26A..52..415M} {52, 415}

\bibitem[\protect\citeauthoryear{{McNamara}, {Kazemzadeh}, {Rafferty},
  {B{\^i}rzan}, {Nulsen}, {Kirkpatrick}  \& {Wise}}{{McNamara}
  et~al.}{2009}]{McNamara2009}
{McNamara} B.~R.,  {Kazemzadeh} F.,  {Rafferty} D.~A.,  {B{\^i}rzan} L.,
  {Nulsen} P.~E.~J.,  {Kirkpatrick} C.~C.,   {Wise} M.~W.,  2009, \mn@doi [ApJ]
  {10.1088/0004-637X/698/1/594}, \href
  {http://adsabs.harvard.edu/abs/2009ApJ...698..594M} {698, 594}

\bibitem[\protect\citeauthoryear{{McNamara}, {Rohanizadegan}  \&
  {Nulsen}}{{McNamara} et~al.}{2011}]{McNamara2011}
{McNamara} B.~R.,  {Rohanizadegan} M.,   {Nulsen} P.~E.~J.,  2011, \mn@doi
  [ApJ] {10.1088/0004-637X/727/1/39}, \href
  {http://adsabs.harvard.edu/abs/2011ApJ...727...39M} {727, 39}

\bibitem[\protect\citeauthoryear{{Mullaney} et~al.,}{{Mullaney}
  et~al.}{2012}]{Mullaney2012}
{Mullaney} J.~R.,  et~al., 2012, \mn@doi [\apjl] {10.1088/2041-8205/753/2/L30},
  \href {http://adsabs.harvard.edu/abs/2012ApJ...753L..30M} {753, L30}

\bibitem[\protect\citeauthoryear{{Nesvadba}, {Lehnert}, {Eisenhauer},
  {Gilbert}, {Tecza}  \& {Abuter}}{{Nesvadba} et~al.}{2006}]{Nesvadba2006}
{Nesvadba} N.~P.~H.,  {Lehnert} M.~D.,  {Eisenhauer} F.,  {Gilbert} A.,
  {Tecza} M.,   {Abuter} R.,  2006, \mn@doi [ApJ] {10.1086/507266}, \href
  {http://adsabs.harvard.edu/abs/2006ApJ...650..693N} {650, 693}

\bibitem[\protect\citeauthoryear{{Nesvadba}, {Lehnert}, {De Breuck}, {Gilbert}
  \& {van Breugel}}{{Nesvadba} et~al.}{2007}]{Nesvadba2007}
{Nesvadba} N.~P.~H.,  {Lehnert} M.~D.,  {De Breuck} C.,  {Gilbert} A.,   {van
  Breugel} W.,  2007, \mn@doi [A\&A] {10.1051/0004-6361:20078175}, \href
  {http://adsabs.harvard.edu/abs/2007A%26A...475..145N} {475, 145}

\bibitem[\protect\citeauthoryear{{Nesvadba}, {Lehnert}, {De Breuck}, {Gilbert}
  \& {van Breugel}}{{Nesvadba} et~al.}{2008}]{Nesvadba2008}
{Nesvadba} N.~P.~H.,  {Lehnert} M.~D.,  {De Breuck} C.,  {Gilbert} A.~M.,
  {van Breugel} W.,  2008, \mn@doi [A\&A] {10.1051/0004-6361:200810346}, \href
  {http://adsabs.harvard.edu/abs/2008A%26A...491..407N} {491, 407}

\bibitem[\protect\citeauthoryear{{Netzer} et~al.,}{{Netzer}
  et~al.}{2007}]{Netzer2007}
{Netzer} H.,  et~al., 2007, \mn@doi [ApJ] {10.1086/520716}, \href
  {http://adsabs.harvard.edu/abs/2007ApJ...666..806N} {666, 806}

\bibitem[\protect\citeauthoryear{{Oliver}, {Bock}, {Altieri}  et~al.}{{Oliver}
  et~al.}{2012}]{Oliver2012}
{Oliver} S.~J.,  {Bock} J.,  {Altieri} B.,   et~al., 2012, \mn@doi [MNRAS]
  {10.1111/j.1365-2966.2012.20912.x}, \href
  {http://adsabs.harvard.edu/abs/2012MNRAS.424.1614O} {424, 1614}

\bibitem[\protect\citeauthoryear{{Oteo}, {Sobral}, {Ivison}, {Smail}, {Best},
  {Cepa}  \& {P{\'e}rez-Garc{\'{\i}}a}}{{Oteo} et~al.}{2015}]{Oteo2015}
{Oteo} I.,  {Sobral} D.,  {Ivison} R.~J.,  {Smail} I.,  {Best} P.~N.,  {Cepa}
  J.,   {P{\'e}rez-Garc{\'{\i}}a} A.~M.,  2015, \mn@doi [\mnras]
  {10.1093/mnras/stv1284}, \href
  {http://adsabs.harvard.edu/abs/2015MNRAS.452.2018O} {452, 2018}

\bibitem[\protect\citeauthoryear{{Puccetti}, {Vignali}, {Cappelluti}
  et~al.}{{Puccetti} et~al.}{2009}]{Puccetti2009}
{Puccetti} S.,  {Vignali} C.,  {Cappelluti} N.,   et~al., 2009, \mn@doi [ApJS]
  {10.1088/0067-0049/185/2/586}, \href
  {http://adsabs.harvard.edu/abs/2009ApJS..185..586P} {185, 586}

\bibitem[\protect\citeauthoryear{{Rodighiero} et~al.,}{{Rodighiero}
  et~al.}{2015}]{Rodighiero2015}
{Rodighiero} G.,  et~al., 2015, \mn@doi [\apjl] {10.1088/2041-8205/800/1/L10},
  \href {http://adsabs.harvard.edu/abs/2015ApJ...800L..10R} {800, L10}

\bibitem[\protect\citeauthoryear{{Rowan-Robinson}}{{Rowan-Robinson}}{1995}]{Rowan-Robinson1995}
{Rowan-Robinson} M.,  1995, MNRAS, \href
  {http://adsabs.harvard.edu/abs/1995MNRAS.272..737R} {272, 737}

\bibitem[\protect\citeauthoryear{{Schaye}, {Crain}, {Bower}  et~al.}{{Schaye}
  et~al.}{2015}]{Eagle2015}
{Schaye} J.,  {Crain} R.~A.,  {Bower} R.~G.,   et~al., 2015, \mn@doi [\mnras]
  {10.1093/mnras/stu2058}, \href
  {http://adsabs.harvard.edu/abs/2015MNRAS.446..521S} {446, 521}

\bibitem[\protect\citeauthoryear{{Schinnerer} et~al.,}{{Schinnerer}
  et~al.}{2004}]{Schinnerer2004}
{Schinnerer} E.,  et~al., 2004, \mn@doi [AJ] {10.1086/424860}, \href
  {http://adsabs.harvard.edu/abs/2004AJ....128.1974S} {128, 1974}

\bibitem[\protect\citeauthoryear{{Schinnerer} et~al.,}{{Schinnerer}
  et~al.}{2007}]{Schinnerer2007}
{Schinnerer} E.,  et~al., 2007, \mn@doi [ApJS] {10.1086/516587}, \href
  {http://adsabs.harvard.edu/abs/2007ApJS..172...46S} {172, 46}

\bibitem[\protect\citeauthoryear{{Schmitt}, {Calzetti}, {Armus}, {Giavalisco},
  {Heckman}, {Kennicutt}, {Leitherer}  \& {Meurer}}{{Schmitt}
  et~al.}{2006}]{Schmitt2006}
{Schmitt} H.~R.,  {Calzetti} D.,  {Armus} L.,  {Giavalisco} M.,  {Heckman}
  T.~M.,  {Kennicutt} Jr. R.~C.,  {Leitherer} C.,   {Meurer} G.~R.,  2006,
  \mn@doi [\apj] {10.1086/501512}, \href
  {http://adsabs.harvard.edu/abs/2006ApJ...643..173S} {643, 173}

\bibitem[\protect\citeauthoryear{{Schnorr M{\"u}ller}, {Storchi-Bergmann},
  {Riffel}, {Ferrari}, {Steiner}, {Axon}  \& {Robinson}}{{Schnorr M{\"u}ller}
  et~al.}{2011}]{Muller2011}
{Schnorr M{\"u}ller} A.,  {Storchi-Bergmann} T.,  {Riffel} R.~A.,  {Ferrari}
  F.,  {Steiner} J.~E.,  {Axon} D.~J.,   {Robinson} A.,  2011, \mn@doi [MNRAS]
  {10.1111/j.1365-2966.2010.18116.x}, \href
  {http://adsabs.harvard.edu/abs/2011MNRAS.413..149S} {413, 149}

\bibitem[\protect\citeauthoryear{{Schweitzer} et~al.,}{{Schweitzer}
  et~al.}{2006}]{Schweitzer2006}
{Schweitzer} M.,  et~al., 2006, \mn@doi [ApJ] {10.1086/506510}, \href
  {http://adsabs.harvard.edu/abs/2006ApJ...649...79S} {649, 79}

\bibitem[\protect\citeauthoryear{{Scoville}, {Aussel}, {Brusa}
  et~al.}{{Scoville} et~al.}{2007}]{Scoville2007}
{Scoville} N.,  {Aussel} H.,  {Brusa} M.,   et~al., 2007, \mn@doi [ApJS]
  {10.1086/516585}, \href {http://adsabs.harvard.edu/abs/2007ApJS..172....1S}
  {172, 1}

\bibitem[\protect\citeauthoryear{{Shankar}, {Weinberg}  \&
  {Miralda-Escud{\'e}}}{{Shankar} et~al.}{2009}]{Shankar2009}
{Shankar} F.,  {Weinberg} D.~H.,   {Miralda-Escud{\'e}} J.,  2009, \mn@doi
  [\apj] {10.1088/0004-637X/690/1/20}, \href
  {http://adsabs.harvard.edu/abs/2009ApJ...690...20S} {690, 20}

\bibitem[\protect\citeauthoryear{{Silk} \& {Rees}}{{Silk} \&
  {Rees}}{1998}]{SilkRees1998}
{Silk} J.,  {Rees} M.~J.,  1998, A\&A, \href
  {http://adsabs.harvard.edu/abs/1998A%26A...331L...1S} {331, L1}

\bibitem[\protect\citeauthoryear{{Sobral} et~al.,}{{Sobral}
  et~al.}{2009a}]{Sobral2009a}
{Sobral} D.,  et~al., 2009a, \mn@doi [MNRAS]
  {10.1111/j.1365-2966.2009.15129.x}, \href
  {http://adsabs.harvard.edu/abs/2009MNRAS.398...75S} {398, 75}

\bibitem[\protect\citeauthoryear{{Sobral} et~al.,}{{Sobral}
  et~al.}{2009b}]{Sobral2009b}
{Sobral} D.,  et~al., 2009b, \mn@doi [MNRAS]
  {10.1111/j.1745-3933.2009.00712.x}, \href
  {http://adsabs.harvard.edu/abs/2009MNRAS.398L..68S} {398, L68}

\bibitem[\protect\citeauthoryear{{Sobral}, {Best}, {Matsuda}, {Smail}, {Geach}
  \& {Cirasuolo}}{{Sobral} et~al.}{2012}]{Sobral2012}
{Sobral} D.,  {Best} P.~N.,  {Matsuda} Y.,  {Smail} I.,  {Geach} J.~E.,
  {Cirasuolo} M.,  2012, \mn@doi [MNRAS] {10.1111/j.1365-2966.2011.19977.x},
  \href {http://adsabs.harvard.edu/abs/2012MNRAS.420.1926S} {420, 1926}

\bibitem[\protect\citeauthoryear{{Sobral}, {Smail}, {Best}, {Geach}, {Matsuda},
  {Stott}, {Cirasuolo}  \& {Kurk}}{{Sobral} et~al.}{2013}]{Sobral2013a}
{Sobral} D.,  {Smail} I.,  {Best} P.~N.,  {Geach} J.~E.,  {Matsuda} Y.,
  {Stott} J.~P.,  {Cirasuolo} M.,   {Kurk} J.,  2013, \mn@doi [MNRAS]
  {10.1093/mnras/sts096}, \href
  {http://adsabs.harvard.edu/abs/2013MNRAS.428.1128S} {428, 1128}

\bibitem[\protect\citeauthoryear{{Sobral}, {Best}, {Smail}, {Mobasher}, {Stott}
   \& {Nisbet}}{{Sobral} et~al.}{2014}]{Sobral2014}
{Sobral} D.,  {Best} P.~N.,  {Smail} I.,  {Mobasher} B.,  {Stott} J.,
  {Nisbet} D.,  2014, \mn@doi [\mnras] {10.1093/mnras/stt2159}, \href
  {http://adsabs.harvard.edu/abs/2014MNRAS.437.3516S} {437, 3516}

\bibitem[\protect\citeauthoryear{{Sobral}, {Kohn}, {Best}, {Smail}, {Harrison},
  {Stott}, {Calhau}  \& {Matthee}}{{Sobral} et~al.}{2016}]{Sobral2016}
{Sobral} D.,  {Kohn} S.~A.,  {Best} P.~N.,  {Smail} I.,  {Harrison} C.~M.,
  {Stott} J.,  {Calhau} J.,   {Matthee} J.,  2016, \mn@doi [\mnras]
  {10.1093/mnras/stw022}, \href
  {http://adsabs.harvard.edu/abs/2016MNRAS.457.1739S} {457, 1739}

\bibitem[\protect\citeauthoryear{{Stanley}, {Harrison}, {Alexander},
  {Swinbank}, {Aird}, {Del Moro}, {Hickox}  \& {Mullaney}}{{Stanley}
  et~al.}{2015}]{Stanley2015}
{Stanley} F.,  {Harrison} C.~M.,  {Alexander} D.~M.,  {Swinbank} A.~M.,  {Aird}
  J.~A.,  {Del Moro} A.,  {Hickox} R.~C.,   {Mullaney} J.~R.,  2015, \mn@doi
  [MNRAS] {10.1093/mnras/stv1678}, \href
  {http://adsabs.harvard.edu/abs/2015MNRAS.453..591S} {453, 591}

\bibitem[\protect\citeauthoryear{{Storchi-Bergmann}, {Lopes}, {McGregor},
  {Riffel}, {Beck}  \& {Martini}}{{Storchi-Bergmann}
  et~al.}{2010}]{Storchi2010}
{Storchi-Bergmann} T.,  {Lopes} R.~D.~S.,  {McGregor} P.~J.,  {Riffel} R.~A.,
  {Beck} T.,   {Martini} P.,  2010, \mn@doi [MNRAS]
  {10.1111/j.1365-2966.2009.15962.x}, \href
  {http://adsabs.harvard.edu/abs/2010MNRAS.402..819S} {402, 819}

\bibitem[\protect\citeauthoryear{{Swinbank}, {Sobral}, {Smail}, {Geach},
  {Best}, {McCarthy}, {Crain}  \& {Theuns}}{{Swinbank}
  et~al.}{2012}]{Swinbank2012}
{Swinbank} A.~M.,  {Sobral} D.,  {Smail} I.,  {Geach} J.~E.,  {Best} P.~N.,
  {McCarthy} I.~G.,  {Crain} R.~A.,   {Theuns} T.,  2012, \mn@doi [\mnras]
  {10.1111/j.1365-2966.2012.21774.x}, \href
  {http://adsabs.harvard.edu/abs/2012MNRAS.426..935S} {426, 935}

\bibitem[\protect\citeauthoryear{{Targett}, {Dunlop}  \& {McLure}}{{Targett}
  et~al.}{2012}]{Targett2012}
{Targett} T.~A.,  {Dunlop} J.~S.,   {McLure} R.~J.,  2012, \mn@doi [\mnras]
  {10.1111/j.1365-2966.2011.20286.x}, \href
  {http://adsabs.harvard.edu/abs/2012MNRAS.420.3621T} {420, 3621}

\bibitem[\protect\citeauthoryear{{Thomson} et~al.,}{{Thomson}
  et~al.}{2014}]{Thomson2014}
{Thomson} A.~P.,  et~al., 2014, \mn@doi [\mnras] {10.1093/mnras/stu839}, \href
  {http://adsabs.harvard.edu/abs/2014MNRAS.442..577T} {442, 577}

\bibitem[\protect\citeauthoryear{{Thomson} et~al.}{{Thomson}
  et~al.}{2016}]{Alasdair}
{Thomson} A.,  et~al., 2016, ApJ, submitted, \href
  {http://adsabs.harvard.edu/abs/2016MNRAS.436.1130S} {}

\bibitem[\protect\citeauthoryear{{Tremonti}, {Moustakas}  \&
  {Diamond-Stanic}}{{Tremonti} et~al.}{2007}]{Tremonti2007}
{Tremonti} C.~A.,  {Moustakas} J.,   {Diamond-Stanic} A.~M.,  2007, \mn@doi
  [ApJL] {10.1086/520083}, \href
  {http://adsabs.harvard.edu/abs/2007ApJ...663L..77T} {663, L77}

\bibitem[\protect\citeauthoryear{{Vasudevan} \& {Fabian}}{{Vasudevan} \&
  {Fabian}}{2007}]{Vasudevan2007}
{Vasudevan} R.~V.,  {Fabian} A.~C.,  2007, \mn@doi [MNRAS]
  {10.1111/j.1365-2966.2007.12328.x}, \href
  {http://adsabs.harvard.edu/abs/2007MNRAS.381.1235V} {381, 1235}

\bibitem[\protect\citeauthoryear{{Young}, {Axon}, {Robinson}, {Hough}  \&
  {Smith}}{{Young} et~al.}{2007}]{Young2007}
{Young} S.,  {Axon} D.~J.,  {Robinson} A.,  {Hough} J.~H.,   {Smith} J.~E.,
  2007, \mn@doi [Nature] {10.1038/nature06319}, \href
  {http://adsabs.harvard.edu/abs/2007Natur.450...74Y} {450, 74}

\bibitem[\protect\citeauthoryear{{Yun}, {Reddy}  \& {Condon}}{{Yun}
  et~al.}{2001}]{Yun2001}
{Yun} M.~S.,  {Reddy} N.~A.,   {Condon} J.~J.,  2001, \mn@doi [ApJ]
  {10.1086/323145}, \href {http://adsabs.harvard.edu/abs/2001ApJ...554..803Y}
  {554, 803}

\makeatother
\end{thebibliography}

\appendix

%
%
%
\section{FIR SED fitting}
\begin{figure*}
	\centering
	\includegraphics[width=16cm]{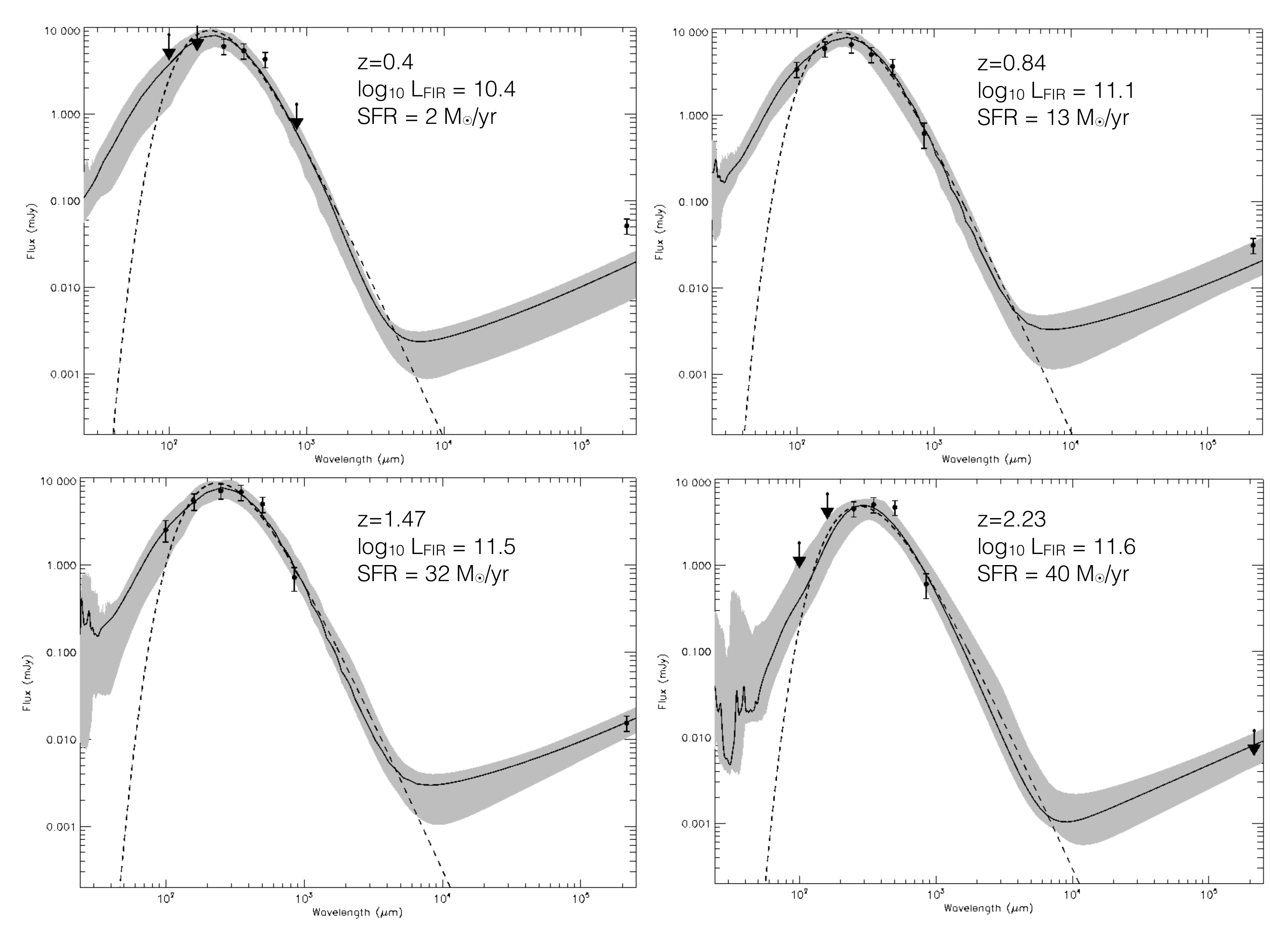}
	\caption{SED fitting for each redshift slice in the far-infrared bands. The data points were obtainned in each band by stacking the entire sample for each redshift using mean statistics. The IR luminosity was estimated by fitting modified black-body templates to the data points and integrating the best fit between 100$\mu$m and 850$\mu$m.}
	\label{fig:SED}
\end{figure*}

\section{Evolution of BHAR and SFR with Stellar Mass}
\begin{figure*}
	\centering
	\includegraphics[width=16cm]{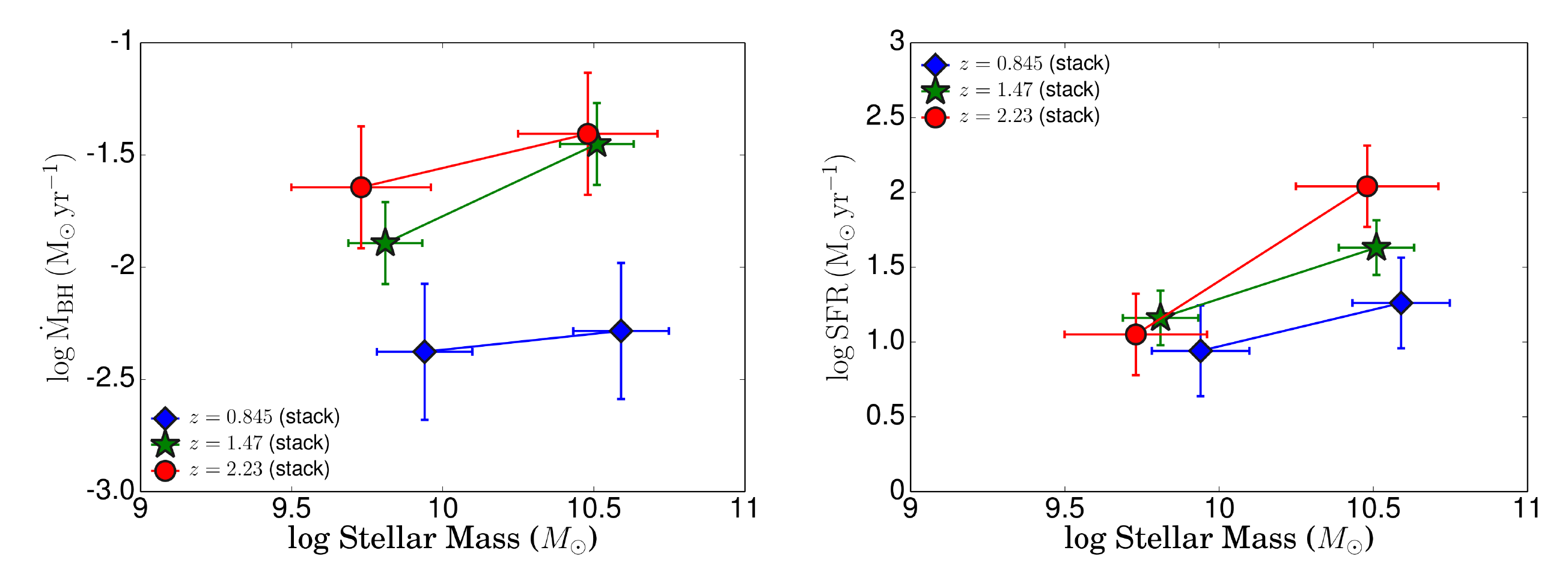}
	\caption{The evolution of the black hole acretion rates and star formation rates with stellar mass for each of the redshift slices in this work. The SFR grows faster than the BHAR, which results in the overal ratio decreasing with stellar mass.}
	\label{fig:SFRBHARMass}
\end{figure*}

\end{document}